\documentclass[10pt]{iopart}
\usepackage[a4paper,margin=1.85cm]{geometry}  

\usepackage{tgtermes}
\usepackage[T1]{fontenc}

\usepackage{graphicx}  

\expandafter\let\csname equation*\endcsname\relax
\expandafter\let\csname endequation*\endcsname\relax
\usepackage{amsmath}
\usepackage{amssymb}  

\usepackage{placeins}

\providecommand{\todo}[1]{}
\newcommand{\dens}[1]{$n_{e,\text{sep}}=#1\cdot10^{19}$\,m$^{-3}$}
\newcommand{\MW}[1]{#1\,MW/m$^2$}
\newcommand{\etaln}{ et al.}
\newcommand{\lined}[1]{$#1\cdot 10^{19}$\,m$^{-2}$}
\newcommand{\Dis}[1]{${D}$ = #1\,m$^\text{2}$/s}
\newcommand{\Dapprox}[1]{${D}$ $\approx$ #1\,m$^\text{2}$/s}
\newcommand{\Dop}[2]{${D} #1 #2$\,m$^\text{2}$/s}
\newcommand{\Dvar}{${D}$}
\newcommand{\chiapprox}[1]{${\chi}$ $\approx$ #1\,m$^\text{2}$/s}
\newcommand{\phiis}[1]{$\varphi\approx #1$\textdegree}
\newcommand{\phiisless}[1]{$\varphi\lessapprox #1$\textdegree}
\newcommand{\phiismore}[1]{$\varphi\gtrapprox #1$\textdegree}
\newcommand{\phirange}[2]{$#1$\textdegree$ \lessapprox\varphi\lessapprox #2$\textdegree}

\graphicspath{{images/}{}}

\providecommand{\bibAnnoteFile}[1]{}
\providecommand{\bibAnnote}[2]{}

\newcommand{\TITLE}{Impact of spatially varying transport coefficients in
  EMC3-Eirene simulations of W7-X and assessment of drifts}
\newcommand{\TITLEL}{\TITLE} 

\providecommand{\AUTHOR}{David Bold}
\providecommand{\AUTHORL}{\AUTHOR}

\newcommand{\baffle}{target}

\usepackage[pdftex,%
	    pdftitle={\TITLEL},%
	    pdfauthor={\AUTHORL},%
	    colorlinks,%
	    citecolor=black,%
	    filecolor=black,%
	    urlcolor=black,%
	    linkcolor=black]{hyperref}
\begin{document}

\title{\TITLE}

\author{  David~Bold${}^{a}$, Felix~Reimold${}^{a}$, Holger~Niemann${}^{a}$, Yu~Gao${}^{a}$,
  Marcin~Jakubowski${}^{a}$, Carsten~Killer${}^{a}$, Victoria~R.~Winters${}^{a}$,
  Nassim~Maaziz${}^{a}$ and the W7-X team${}^{b}$}

\address{
  ${}^{a}$Max Planck Institute for Plasma Physics, Wendelsteinstr. 1, 17491
  Greifswald, Germany\\
  \todo{update below}
  ${}^{b}$See Pedersen et al 2022~\cite{pedersen22a} for the W7-X Team
  }
\ead{dave@ipp.mpg.de}
\vspace{10pt}
\begin{indented}
\item[]\today
\end{indented}


\begin{abstract}
  Modelling the scrape-off layer of a stellarator is challenging due
  to the complex magnetic 3D geometry.  The here presented study
  analyses simulations of the scrape-off layer (SOL) of the
  stellarator Wendelstein 7-X (\mbox{W7-X}) using spatially varying diffusion coefficients for
  the magnetic standard configuration, extending our previous
  study~\cite{bold22a}.  Comparing the EMC3-Eirene simulations with
  experimental observations, an inconsistency between the strike-line width (SLW) and the upstream
  parameters was observed. While to match the experimental SLW
  a particle diffusion coefficient \Dapprox{0.2} is needed, 
  \Dapprox{1} is needed to get experimental separatrix temperatures of 50\,eV
  at the given experimental heating power.  We asses the impact of physically
  motivated spatially varying transport coeffients.  Agreement with experimental data
  can be improved, but various differences remain.  We show that drifts are expected to
  help overcome the discrepancies and, thus, the development of SOL
  transport models including drifts is a necessary next step to study
  the SOL transport of the \mbox{W7-X} stellarator.
\end{abstract}

\FloatBarrier
\section{Introduction}\label{s:intro}
In order to operate fusion power plants based on the magnetic confinement
concept the power flux on the plasma-facing surfaces needs to be controlled to
prevent the overloading of structures.
Predictive modelling of power loads, with high fidelity codes, such as
EMC3-Eirene, will be necessary for the design of a next-step fusion device.
Successful validation via comparison to existing experimental devices is
required to
ensure all important underlying physics is included in the code.
One of these devices where we can do such a validation is Wendelstein 7-X (\mbox{W7-X}), an
optimised stellarator with reduced neoclassical
transport~\cite{pedersen22a,klinger17a,wolf17a,pedersen17a,beidler21a}.

\begin{figure}
  \centering
  \raisebox{-0.5\height}{\includegraphics[width=.3\linewidth]{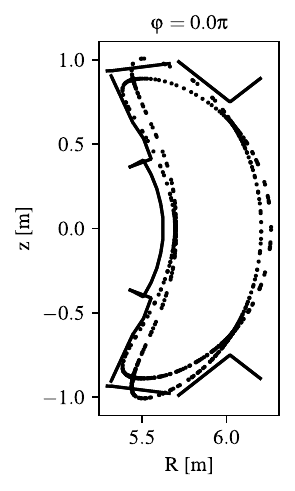}}
  \raisebox{-0.5\height}{\includegraphics[width=.4\linewidth]{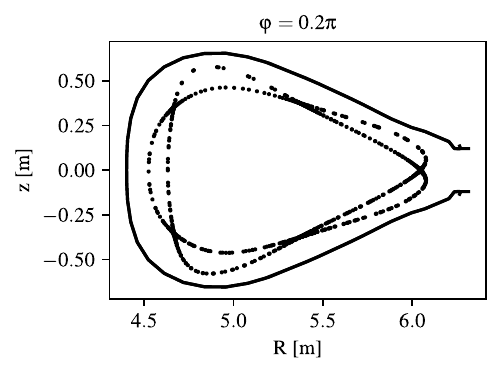}}
  \caption{Shown are the islands at toroidal position $\varphi = 0$ and
    $\varphi=\pi/5 = 36$\,\textdegree{} as dots as well as the target structures
    used in the simulations, reproduced with kind permissions from~\cite{bold22a}.
  }\label{f:poincare}
\end{figure}
\begin{figure}
  \centering
  \includegraphics[width=.7\linewidth]{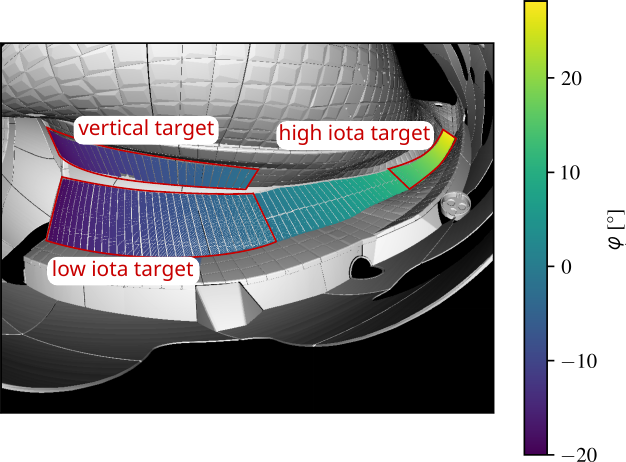}
  \caption{View into the divertor as seen from the IR camera. The divertor
    targets are shown colour coded. The colour represent the toroidal
    angle. The main, horizontal target is the larger of the two structures at
    the bottom. The vertical target is on the top, and only ranges to
    $-20$\textdegree~$ < \varphi < 0$. In total 137 target structures, so
    called fingers, are visible, that have a toroidal extend of around
    $0.5$\textdegree.}\label{f:fingerpos}
\end{figure}

In contrast to tokamaks, the scrape-off layer (SOL) of \mbox{W7-X} is inherently
three dimensional.  \mbox{W7-X} features a 5-fold toroidal symmetry. Each of the
5 modules is in itself stellarator symmetric and can be split into two half
modules.
The SOL of \mbox{W7-X} features an island divertor, where in the
standard configuration the 5 resonant islands are intersected by 10
divertor modules~\cite{pedersen18a,pedersen19a,hammond19a}. A plot of the
islands and the intersection with the divertor is shown in
fig.~\ref{f:poincare}. The view of the thermography used later in this paper
on to the divertor is shown in fig.~\ref{f:fingerpos}.

The lack of toroidal symmetry makes the comparison of
experimental measurements at different toroidal locations challenging.
As a consequence, there is great need for 3D modelling, where
synthetic diagnostics can be implemented to help understand whether
differing diagnostic measurements are truly in disagreement or if
differences are due purely to spatial variations in the
plasma that are consistent with the implemented physics model. However, before
such an analysis can be performed, it is
critical to first validate the simulations, which itself requires
diagnostic input covering as much of the SOL plasma domain as
possible.

The anomalous cross-field transport in the SOL of fusion plasmas is
often considered to be dominated by turbulence. In \mbox{W7-X} experiments,
SOL turbulence and turbulent transport have been
observed~\cite{killer20a,killer21b,zoletnik19a,liu19a}.
Fully turbulent simulations of the
full SOL are computationally extremely challenging and not availaible for 3D
stellarator geometries yet. Simpler models
are generally used, such as fluid transport codes~\cite{winters21a}.
There the turbulence is effectively represented by anomalous diffusion
coefficients. The
simplest diffusion model features a spatially constant diffusion
coefficient. Spatially varying diffusion coefficients give significantly more
freedom in matching data and have been used in tokamak simulations in the
past~\cite{lunt20a,zhang16a,carli20a}.  However, finding appropriate
distributions of the diffusion coefficient is challenging due to the large
parameter space. To limit the parameter space, the coefficients can,
for example, be motivated by experimental observations,
theoretical predictions, or turbulence simulations~\cite{carli20a}.

This work validates the diffusion-based anomalous transport in the absence of drifts in the
model in EMC3-Eirene~\cite{feng14a,feng21a} by extending the isotropic
diffusion scan~\cite{bold22a} to spatially varying transport
coefficients. The simulations are compared to experimental data
from \mbox{W7-X}'s infra-red heat-flux diagnostic and the reciprocating
electric probes.
\todo{TS?}
\todo{He-beam?}
In particular, the radial profiles of electron temperature and density, the
strike-line width and toroidal distribution of heat-flux onto the divertor
will be compared.  The analysis presented here is restricted to the magnetic
standard configuration.

The scope of the paper is to asses the impact of spatially varying and
physically motivated transport coefficients rather than exhaustive variations
to fit experimental data. The general impact of the applied variations is
discussed to elucidate the role of the variation.

This paper is organized as follows: Section~\ref{s:method} 
the diffusion coefficient profiles are
introduced and motivated.
In section~\ref{s:exp} the experimental conditions are summarized, a detailed
description is given in ref.~\cite{bold22a}.
In section~\ref{s:simulations}, the simulations are presented.
Section~\ref{s:discussion} summarises and discusses the results, here
it is seen that a spatially varying diffusion coefficient can
improve matching downstream and upstream conditions in the
magnetic standard configuration.
The main conclusions are presented in the final section.

\section{Method}\label{s:method}
\subsection{\mbox{W7-X} diagnostics}\label{s:diag}
As in previous work~\cite{bold22a}, two diagnostics are used for comparison to simulation
data: one is located downstream at the divertor targets and the other
upstream. Both downstream and upstream parameter comparisons are
important to determine if the EMC3-Eirene simulations successfully
reproduce features across the entire SOL.  The downstream measurement
used is the infra-red (IR) camera system~\cite{jakubowski18a}, which
fully covered the area of 9 out of 10 divertors in the previous
experimental campaign OP 1.2b. The view onto the target is shown in
fig.~\ref{f:fingerpos}.

The temperature is derived from the spatial distribution of the IR
radiation. The heat-flux is calculated by the evolution of the temperature
profiles using the two-dimensional thermal model THEODOR~\cite{sieglin15a}.
This heat-flux is used to assess the validity of the employed SOL
transport model in EMC3-Eirene as the heat-flux has a high spatial
resolution. The spatial resolution is around 3\,mm and the noise level is
around \MW{0.25}.

The second diagnostic used for comparison are reciprocating Langmuir probes
mounted on the Multi-Purpose Manipulator (MPM) which provide profiles of
electron temperature and density~\cite{killer19a,killer21a}.
\begin{figure}
  \centering
  \includegraphics[width=.7\linewidth]{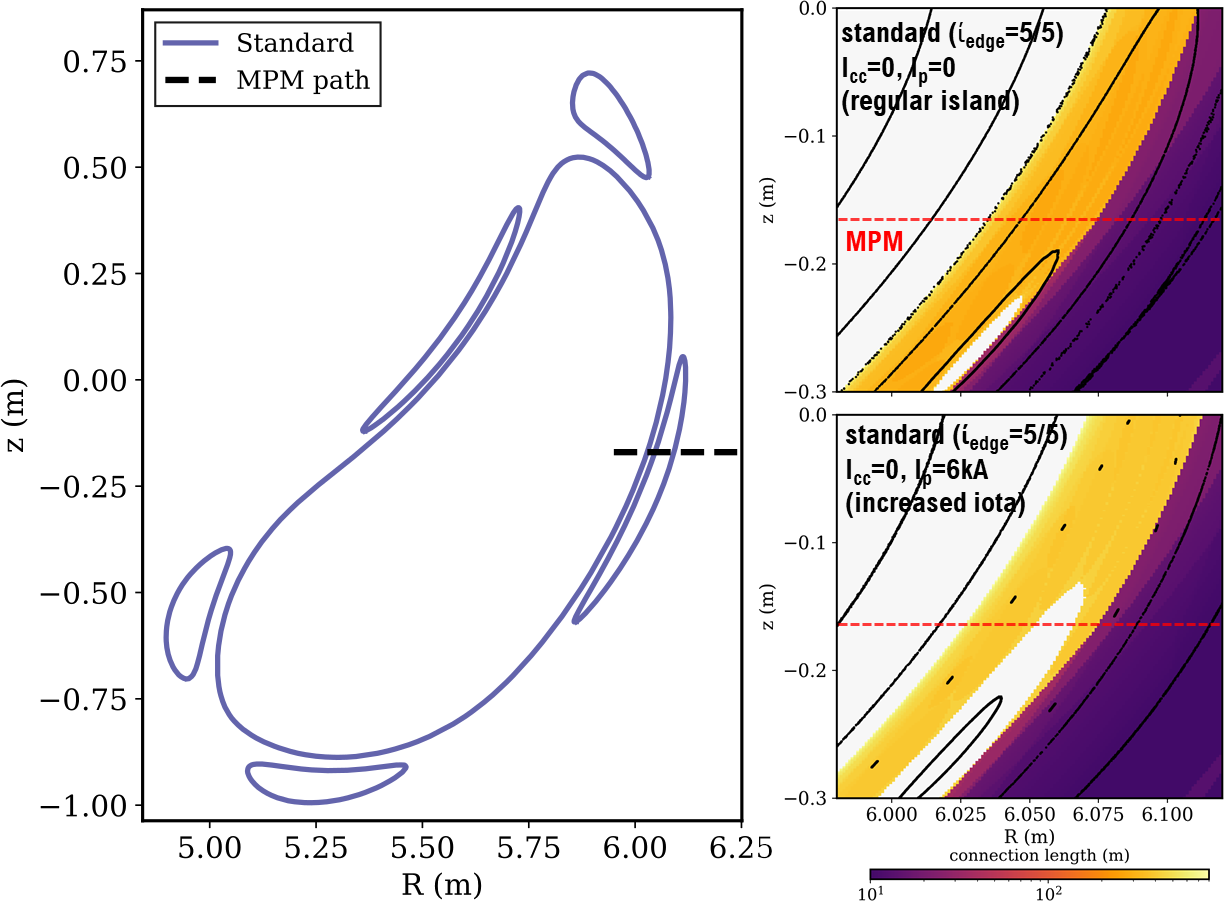}
  \caption{Position of the MPM diagnostic, showing the islands and last-closed
    flux surface (LCFS) on the left, and a plot of the connection length on
    the right, for the ideal case of no toroidal plasma current, as assumed in
    the simulations, on the top and for a finite plasma current of 6\,kA on
    the bottom. Reproduced with kind permission from~\cite{killer21a}.
  }\label{f:mpm:pos}
\end{figure}
This provides plasma parameters
upstream and, thus, complements the downstream comparison provided by
the heat-flux measurements at the divertor. Unlike the infra-red
diagnostic the MPM is only present in one half module and, therefore, does not give
a direct measurement of up-down asymmetries~\cite{hammond19a} or field
errors~\cite{lazerson18a}.
The path of the MPM is shown in fig.~\ref{f:mpm:pos}.

It is crucial to include measurments from different locations, to constrain
the model. Otherwise (de)validation of a model is not possible. Especially
with a raising number of degrees of freedom in the model, sufficient amount of
experimental data needs to be used for comparison.

\subsection{Heat-Flux distribution analysis}\label{s:fluxana}
The strike-line width and amplitude are used to compare the
heat-flux profiles between experiment and modelling.
For all toroidal locations of each target,
the poloidal position, amplitude and strike-line width is determined, by a fit
of
the main heat-flux peak along the strike-line.
For a detailed description of the mapping and analysis of the
heat-flux, see our previous work~\cite{bold22a}.


\begin{figure}
  \centering
  \includegraphics[width=.49\linewidth]{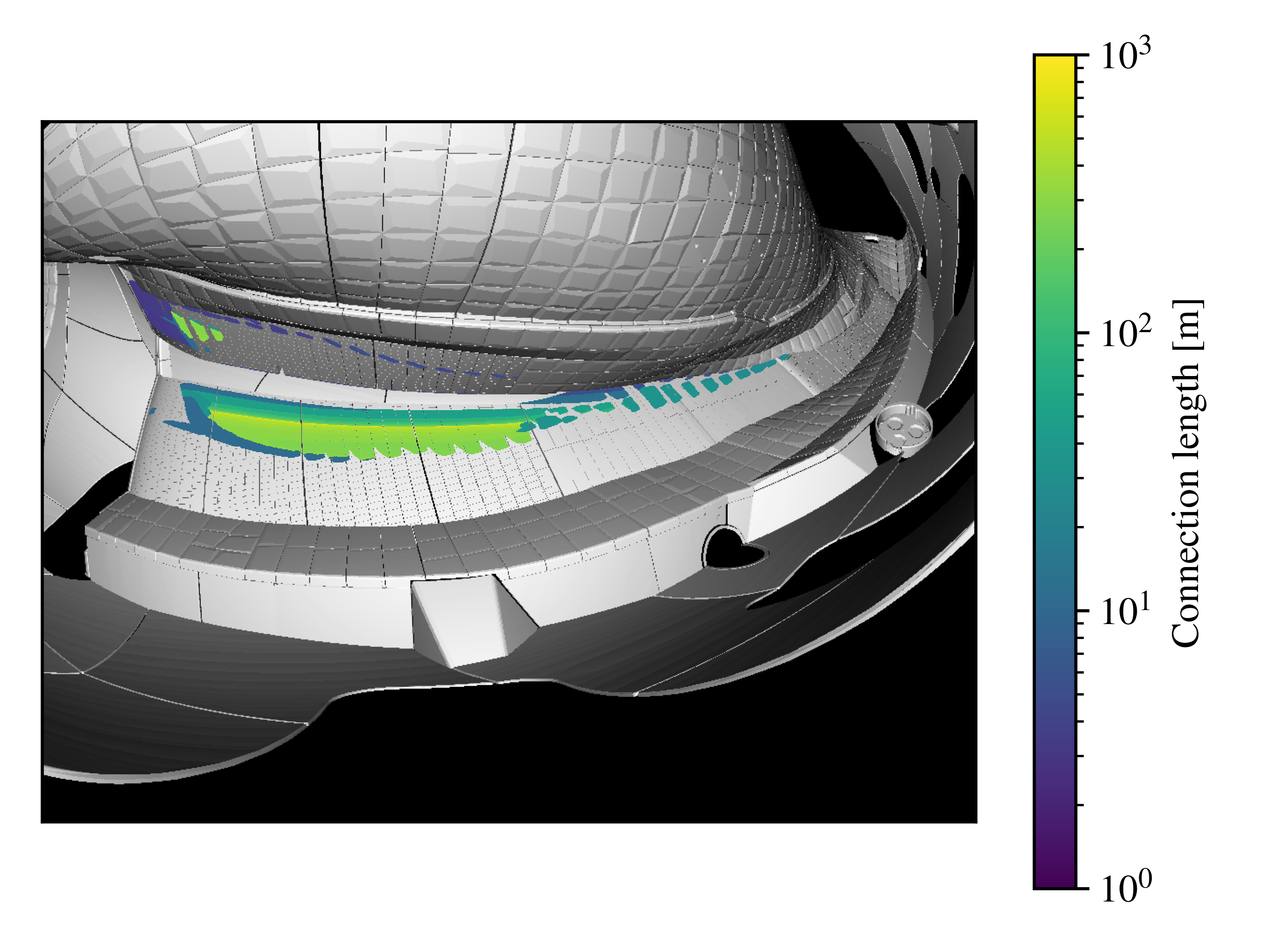}
  \includegraphics[width=.49\linewidth]{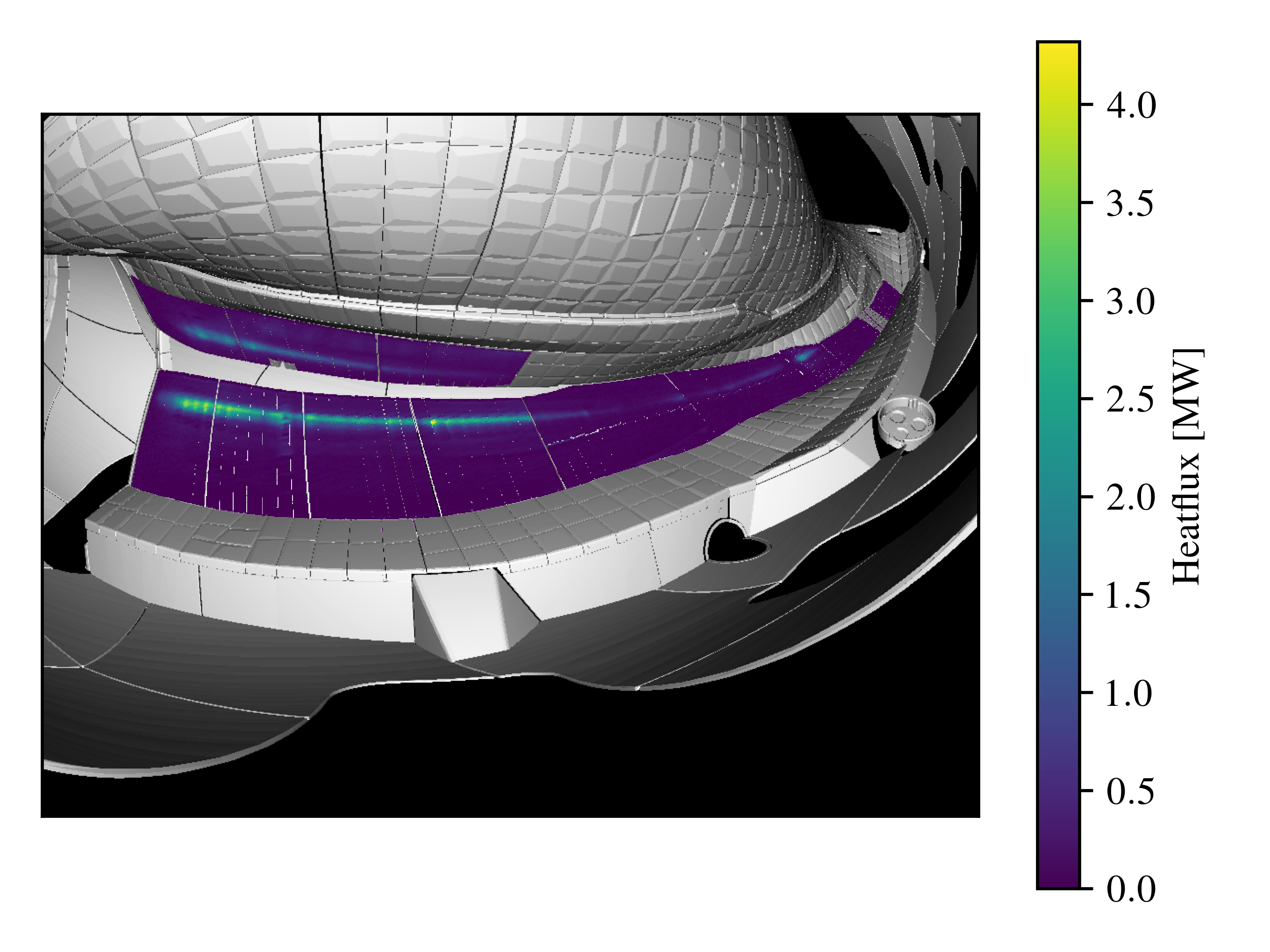}
  \caption{Shown on the left is a plot of the connection length mapped
    on to the target.  In grey
    the target regions where no traced field line ended are shown.
    On the right is a  plot of the heat-flux on the divertor at
    $t=t_1+3.3$\,s for shot \#20180920.009.
    The main strike-line is on the horizontal target \phiisless{0},
    roughly in
    agreement with the long connection lengths. Additional
    heat loads on the horizontal target around \phiis{13} as
    well as the vertical \baffle{} are visible.
  }\label{f:con}
\end{figure}

\begin{figure}
  \centering
  \includegraphics[width=.49\linewidth]{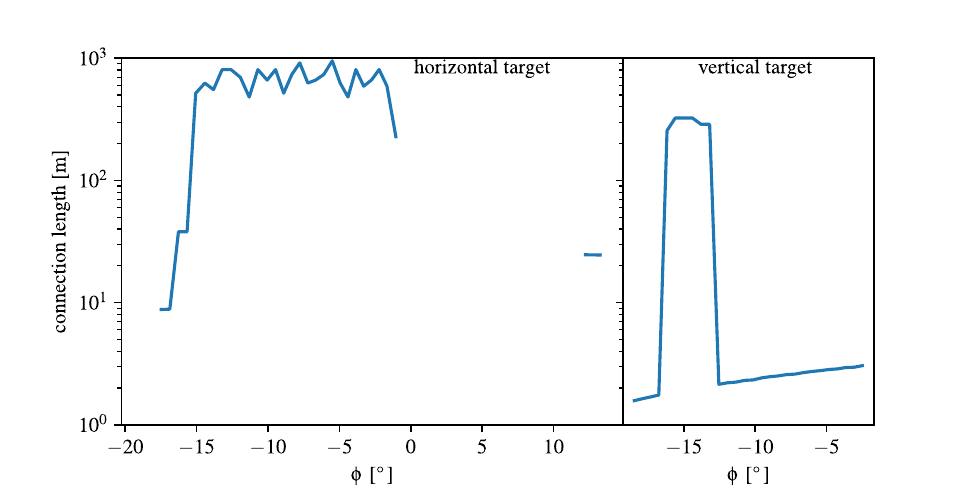}
  \caption{The maximum of connection length as shown in fig
    \ref{f:con} as a function of toroidal angle.
  }\label{f:conpar}
\end{figure}

Fig.~\ref{f:con} shows the connection length ($L_c$) mapped to the
target regions along with a typical heat load pattern.
Fig.~\ref{f:conpar} shows the maximum $L_c$ as a
function of toroidal angle for the horizontal and vertical target, see fig.~\ref{f:fingerpos}.  Regions of very long connection length $> 500\,$m
indicate the location of the main strike-line formed by the
intersection of the edge of the island on the divertor target plates.
For the standard configuration, the main strike-line is on the horizontal
target, but,
also on the vertical \baffle{} long connection lengths are observed.

\subsection{EMC3-Eirene}\label{s:emc3}
EMC3-Eirene has been used to simulate the scrape-off layer in the current work.
EMC3-Eirene is a Monte Carlo fluid transport and kinetic neutral code, that is capable of
handling complex geometries, such as those commonly encountered in the SOL of
stellarators. It has already been used in the past to model the edge of
\mbox{W7-X}~\cite{feng21a,winters21a,bold22a}. While EMC3-Eirene does captures some of the
observations in experiments, especially global
trends~\cite{winters21a,feng21a}, there is still disagreement in local
parameters~\cite{feng21b,bold22a}.

EMC3-Eirene does include parallel transport in the form of advection as well
as viscosity and parallel heat diffusivity.  Perpendicular transport included
in EMC3-Eirene features anomalous diffusion based on some given particle and
heat diffusion coefficients, that can be spatially
varying~\cite{feng14a,zhang16a}. EMC3-Eirene does not require nested
flux-surfaces and is only ``aware'' of the local magnetic geometry, i.e.~it is
ignorant of island flux surfaces.  For this
reason the perpendicular diffusion is uniform in flux-surface perpendicular and bi-normal
directio.
Drifts
are not included in EMC3-Eirene.

The simulations are analysed using the same procedure as described
in~\cite{bold22a}.
Both experimental as well as simulation data have been fitted using the same
routines to produce a
more consistent data set for comparison.

\section{Experimental data}\label{s:exp}
Despite the qualitative modelling approach, the analysis should be
done in realistic conditions and with reference to experimental data.
The \mbox{W7-X} experiments \#20180920.009 and
\#20180920.013 
have been analysed and are compared to the simulations. They are part
of a density scan with an input power of 4.7\,MW ECRH.
For a more detailed description of the
experiments, see our previous work~\cite{bold22a}.

The line integrated density was \lined{4} and 
\lined{6}. These low and medium density cases were selected as they feature a
low radiative fraction $f_{rad} = P_{rad}/P_{heat}$ of 0.15 to 0.35.
This allows us to focus on the effect of heat transport on
the target heat load distribution,
reducing the additional impact of radiation as volumetric loss.

The data for the MPM comes the experiments \#20181010.008, \#20181010.021,
\#20181010.022 and \#20181010.016. They are similar to the other two
experiments and then density is in the range of \lined{4.5\ldots6}.

\section{Simulations}\label{s:simulations}
In order to decrease the discrepancies observed in the previous
study~\cite{bold22a}, where low values of the diffusion coeffients where
needed to match the strike-line with, and high values where needed to reduce
the upstream temperature, spatially varying diffusion coeffients are assesed
here for their potential to reconcile the upstream and downstream measurements.
These scenarios are
not intended to model the SOL of \mbox{W7-X} in a realistic way, but rather to
measure the impact of the changes onto the synthetic diagnostics,
test their ability to improve agreement, and elucidate what transport aspects
are changed or not.
Another question addressed here is whether the different transport models
could be distinguished by diagnostics in experiments via specific features or
scalings.

\subsection{Spatially varying diffusion}\label{s:m:dist}
\begin{figure}
  \centering
  \raisebox{-0.5\height}{\includegraphics[width=.49\linewidth]{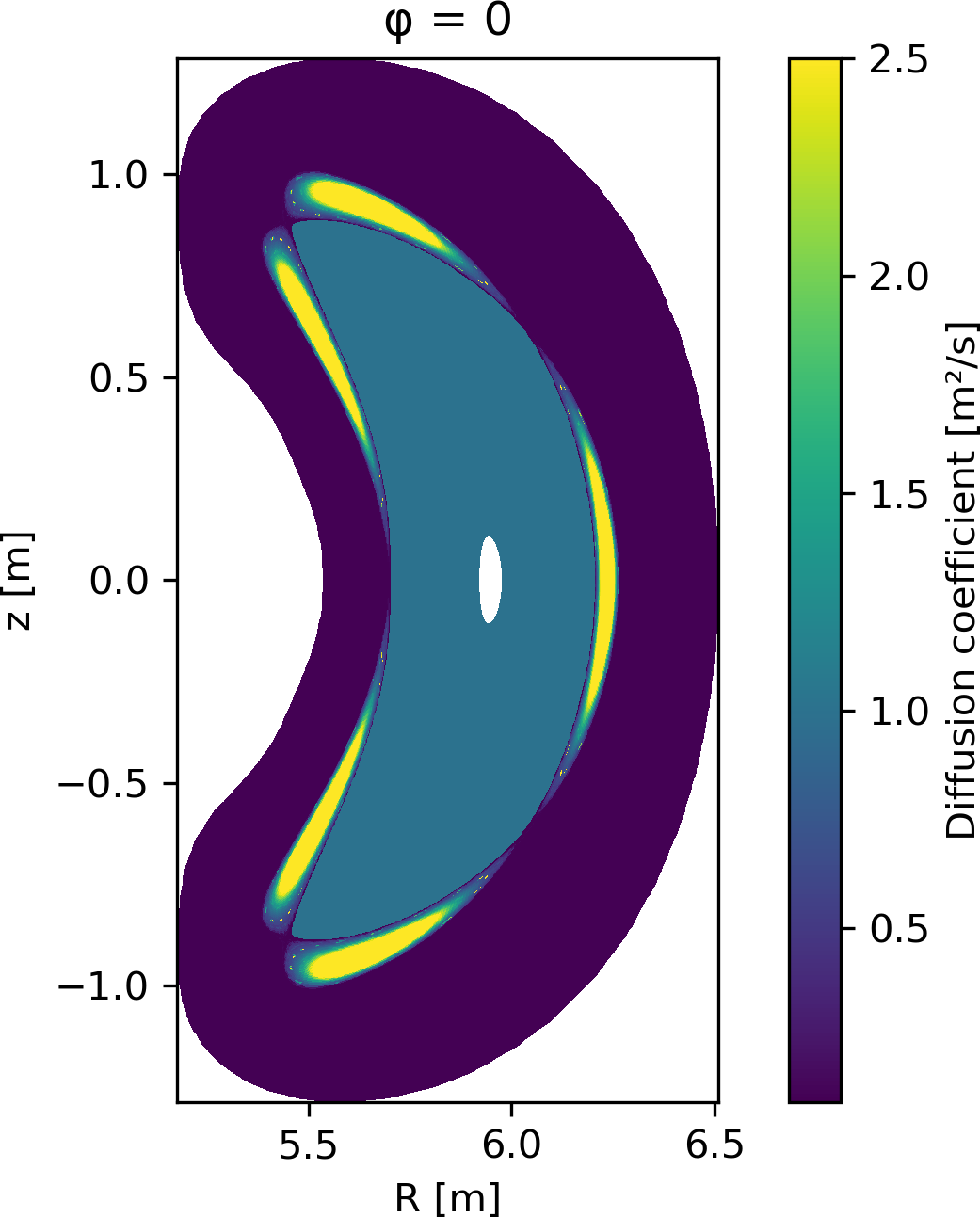}}
  \raisebox{-0\height}{\begin{minipage}{.49\linewidth}%
      \includegraphics[width=\linewidth]{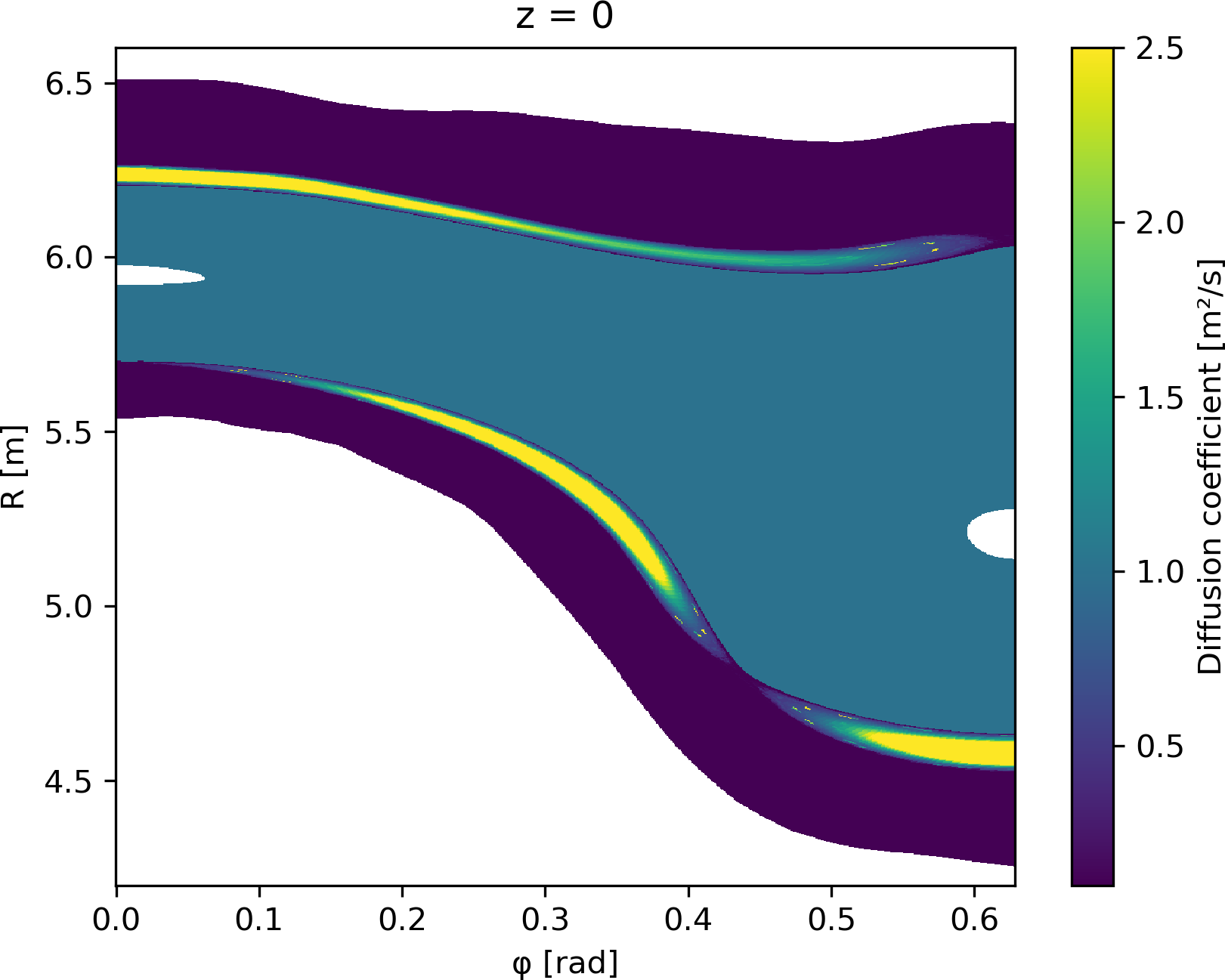}\\%
      \includegraphics[width=\linewidth]{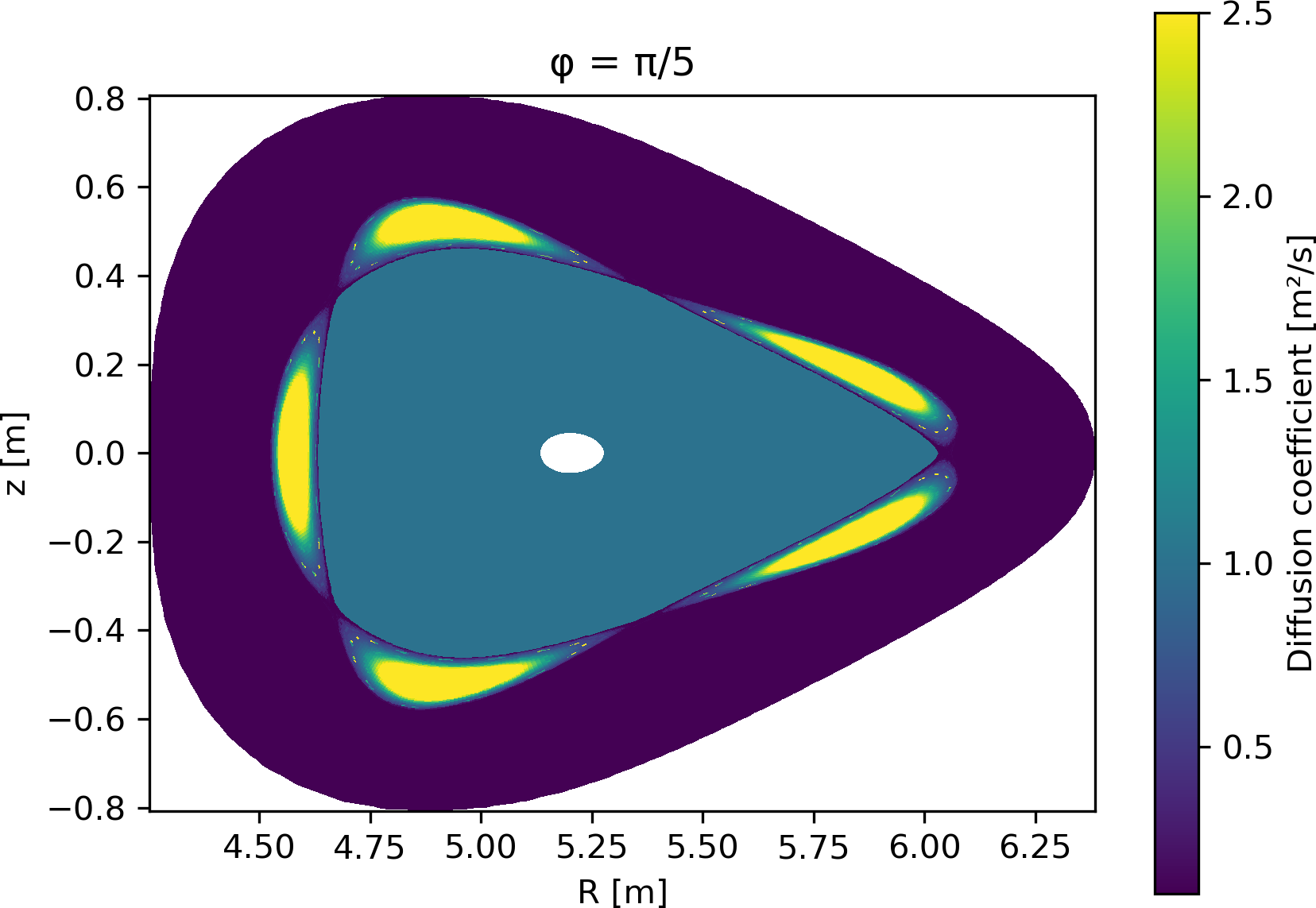}%
  \end{minipage}}
  \caption{Plot of scenario~B: diffusion based on experimental data featuring
    radial-poloidal variation.  On the left is a cut at $\varphi=0$, on the
    right top at $z=0$ and on the right bottom $\varphi=\pi/5=36$\textdegree.  The diffusion is set to \Dis{1} in the core region for
    numerical stabilities of the boundary conditions. Further outside a
    background of \Dis{0.1} is used, and enhanced towards the centre of the
    the islands.
  }\label{f:D:isl}
\end{figure}

\begin{figure}
  \centering
  \raisebox{-0.5\height}{\includegraphics[width=.49\linewidth]{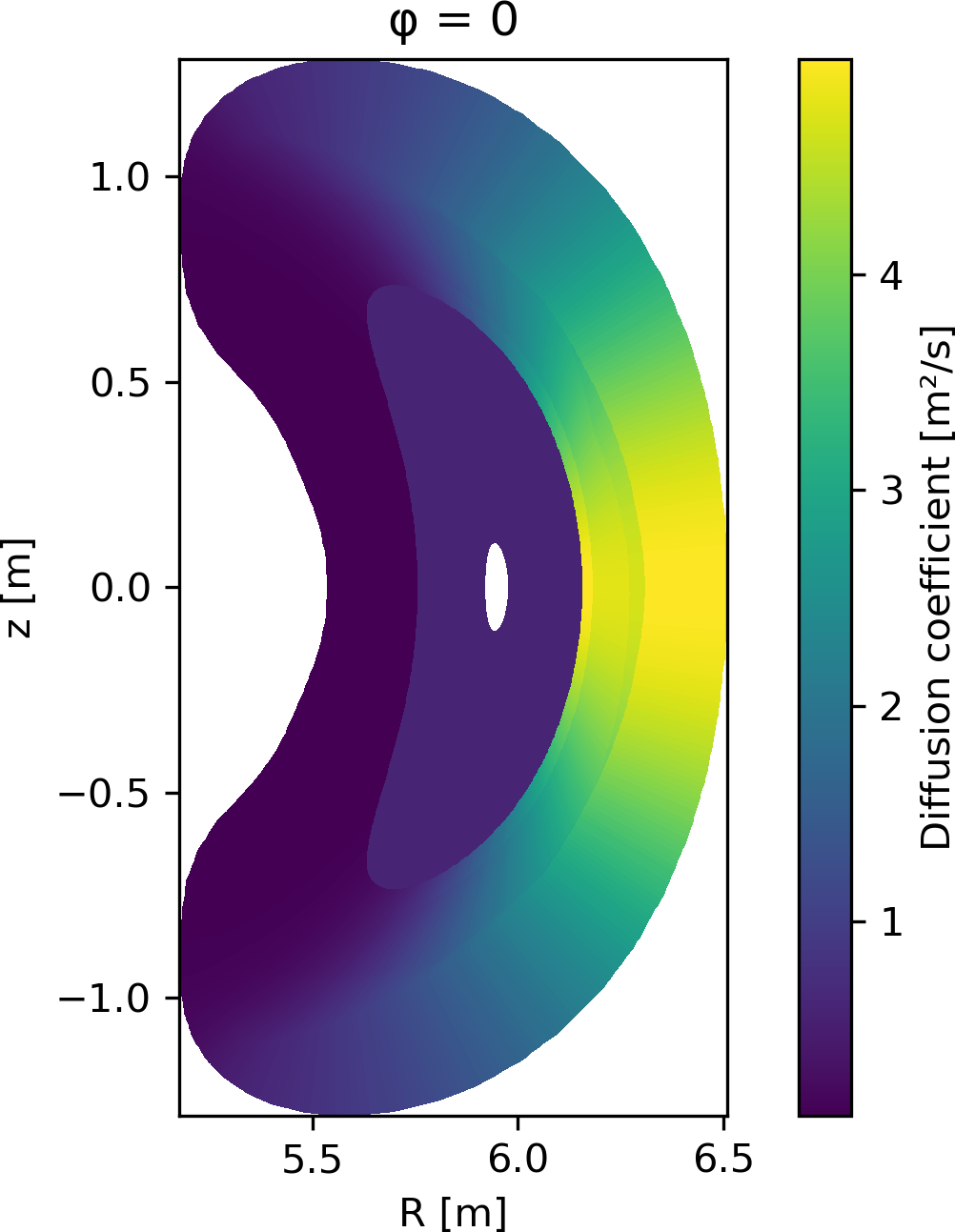}}
  \raisebox{+0\height}{\begin{minipage}{.49\linewidth}\includegraphics[width=\linewidth]{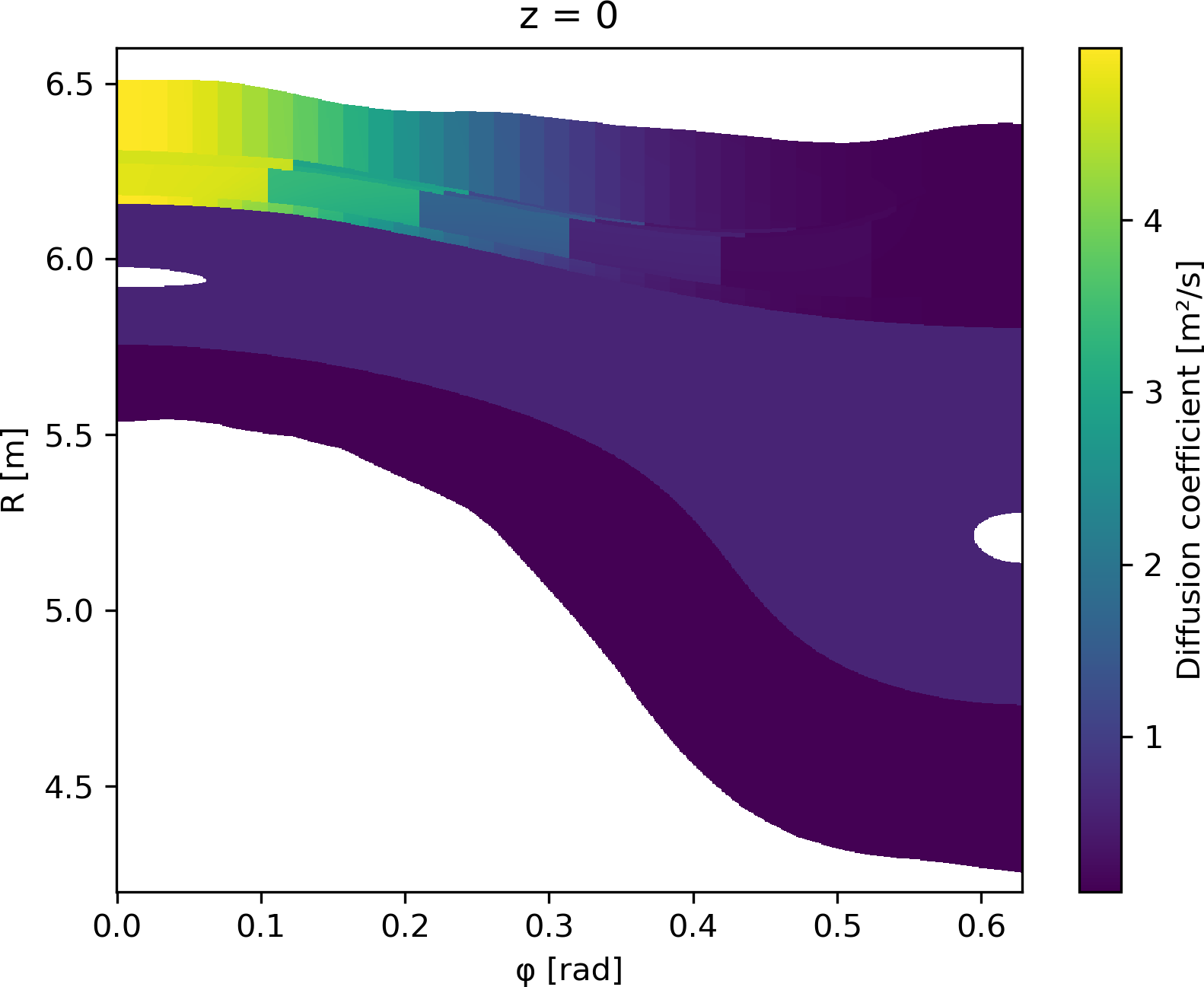}\\
      \includegraphics[width=\linewidth]{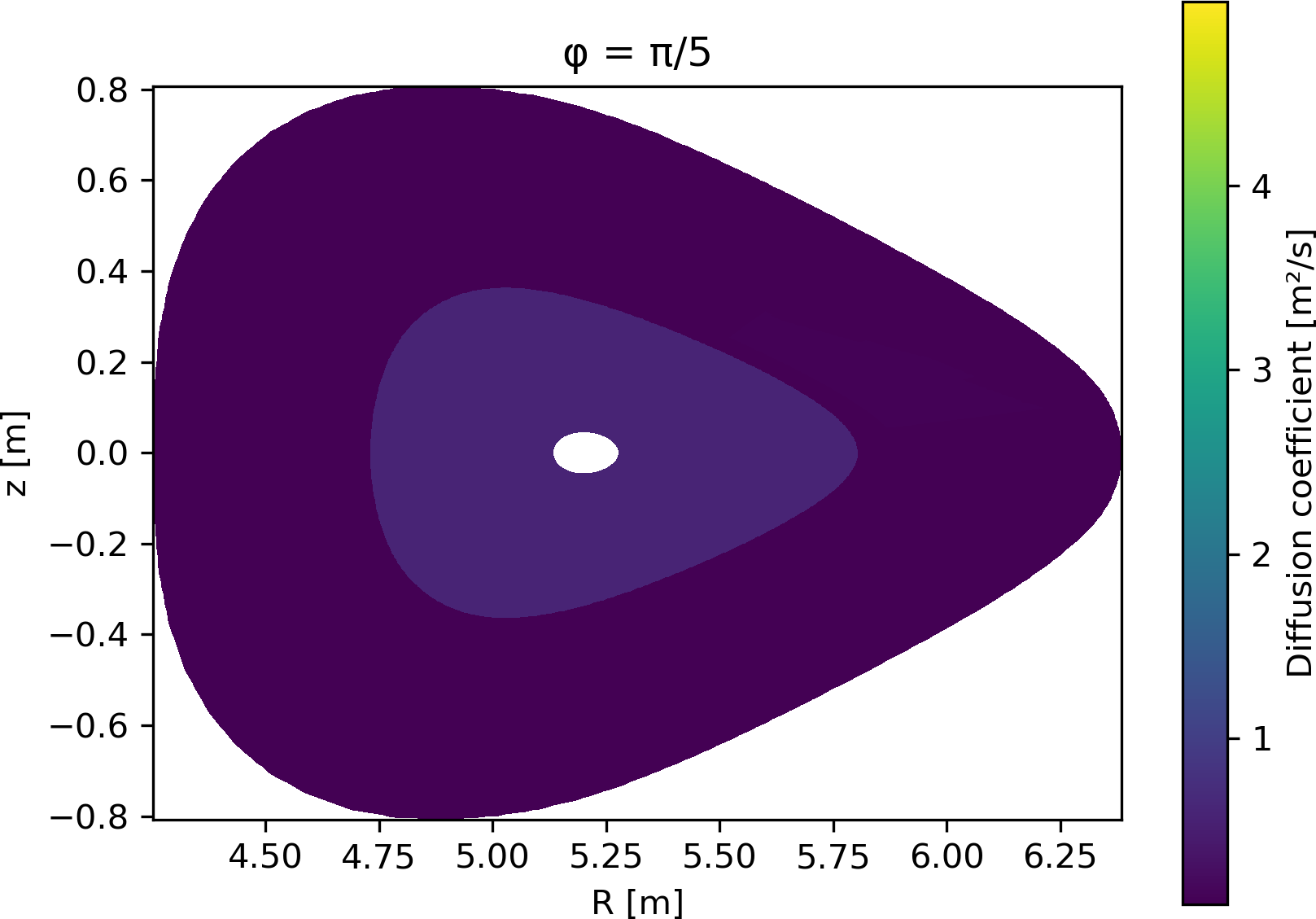}
  \end{minipage}}
  \caption{Plot of scenario~C: diffusion based on turbulence metrics featuring
    a poloidal-toroidal variation. On the
    left is a cut at $\varphi=0$, on the
    right top at $z=0$ and on the right bottom
    $\varphi=\pi/5=36$\textdegree. The diffusion is enhanced
    in the core region for numerical stabilities of the boundary
    conditions
    Further outside a background of \Dis{0.1} is used and enhanced
    in the outer bean shape.
  }\label{f:D:poltor}
\end{figure}


The simulations presented here use spatially varying heat and particle diffusion
coefficients, which are implemented in EMC3-Eirene.
In addition to scenario~A: constant diffusion presented in ref.~\cite{bold22a},
two spatial variation patterns were implemented in this
work:  
scenario~B, motivated by experimental observations and scenario~C,
motivated by turbulence characteristics. Scenario~B is shown in fig.~\ref{f:D:isl}.
The transport is suppressed at the separatrix and enhanced towards
the island centres.  The distribution is motivated by experimentally
observed heat-fluxes, as the narrow strike-line requires a low transport coefficient at
the separatrix, but for agreement with other diagnostics measuring away from
the strike-line, such as the
MPM, higher transport coefficients are required~\cite{feng21a}.  Thus,
combining a low perpendicular transport value at the separatrix and
a higher value towards the centre of the island can better satisfy both
conditions.
It will lead to a tightly confined power carrying layer, focusing the parallel heat-flux
towards the divertor while broadening the profiles upstream towards the O-point.
Fig.~\ref{f:D:isl} on the right shows the coefficients at
$z=0$. The variation along the toroidal direction $\varphi$ in fig.~\ref{f:D:isl} $z=0$ is due to the
poloidal contribution of the magnetic field, as the variation of $D$ is
aligned with the magnetic field and the island rotates around the LCFS.

Scenario~C, motivated by
turbulent transport drivers is shown in
fig.~\ref{f:D:poltor}. In this scenario the transport is enhanced in the outer
bean shape, where a Gaussian perturbation has been added to a constant background. The outer
bean shape features the largest values of bad curvature, which is a
significant driver for turbulent transport in the
SOL~\cite{killer20a,schworer18a}.
Thus, this set of simulations mimics in a naïve way the effect of turbulent
transport and probes the impact of toroidally localised power flux
into the SOL, similar to balloning type assumptions in tokamak modelling.

In all cases the heat diffusion coefficient $\chi$ is set to
$\chi = 3\cdot D$, i.e. scaled with the particle diffusion coefficient.
Note that the
resulting transport $q_\perp \propto n \chi$, i.e. has a density
dependence even for constant $\chi$.

\subsection{Simulation parameter}
The scrape-off layer of \mbox{W7-X} has been modelled using
EMC3-Eirene. For this the upstream density was scanned.
The simulation relies on the stellarator symmetry of \mbox{W7-X}'s standard geometry and,
therefore, only one half module is modelled. Ideal coils are used and,
thus, no error field effects are included. Drifts are not included.

Input parameters are set consistently with~\cite{bold22a}. The input heating power
within the simulation domain (one half module)
was set to be 470\,kW, leading to a total of 4.7\,MW for the whole
device.
The power was
distributed evenly between ions and electrons, and enters the domain at the
core boundary. The observed power on the divertor is up to 352\,kW -
giving a total power of $\approx 3.5$\,MW on all divertors.
The radiation was fixed to 1\,MW, achieved via carbon impurity radiation, giving a radiation
fraction $f_{rad} \approx 21$\,\%. In the experiment the radiation fraction
varies from $15$\,\% to $35$\,\%.
The low $f_{rad}$ was selected to
avoid a dominant effect of the radiation losses.

The upstream density was set to a fixed value of \dens{1} and \dens{3}.
The cases are roughly in the range of the experiments.
No
pumping and fuelling is included in the simulations and, therefore,
particle balance is achieved via scaling of the recycling flux to the
amount needed for the fixed upstream density value.

The same magnetic field configuration was used in simulation as in
experiment: the standard magnetic field configuration, featuring a 5/5 island
chain in the SOL.

For each diffusion coefficient setup described in
sec.~\ref{s:m:dist}, a scan in density and in magnitude of diffusion coefficient was
performed. The case with uniform diffusion coefficients (scenario~A) is also
included for reference.

\begin{figure}
  \centering
  \raisebox{-0.5\height}{\includegraphics[width=.4\linewidth]{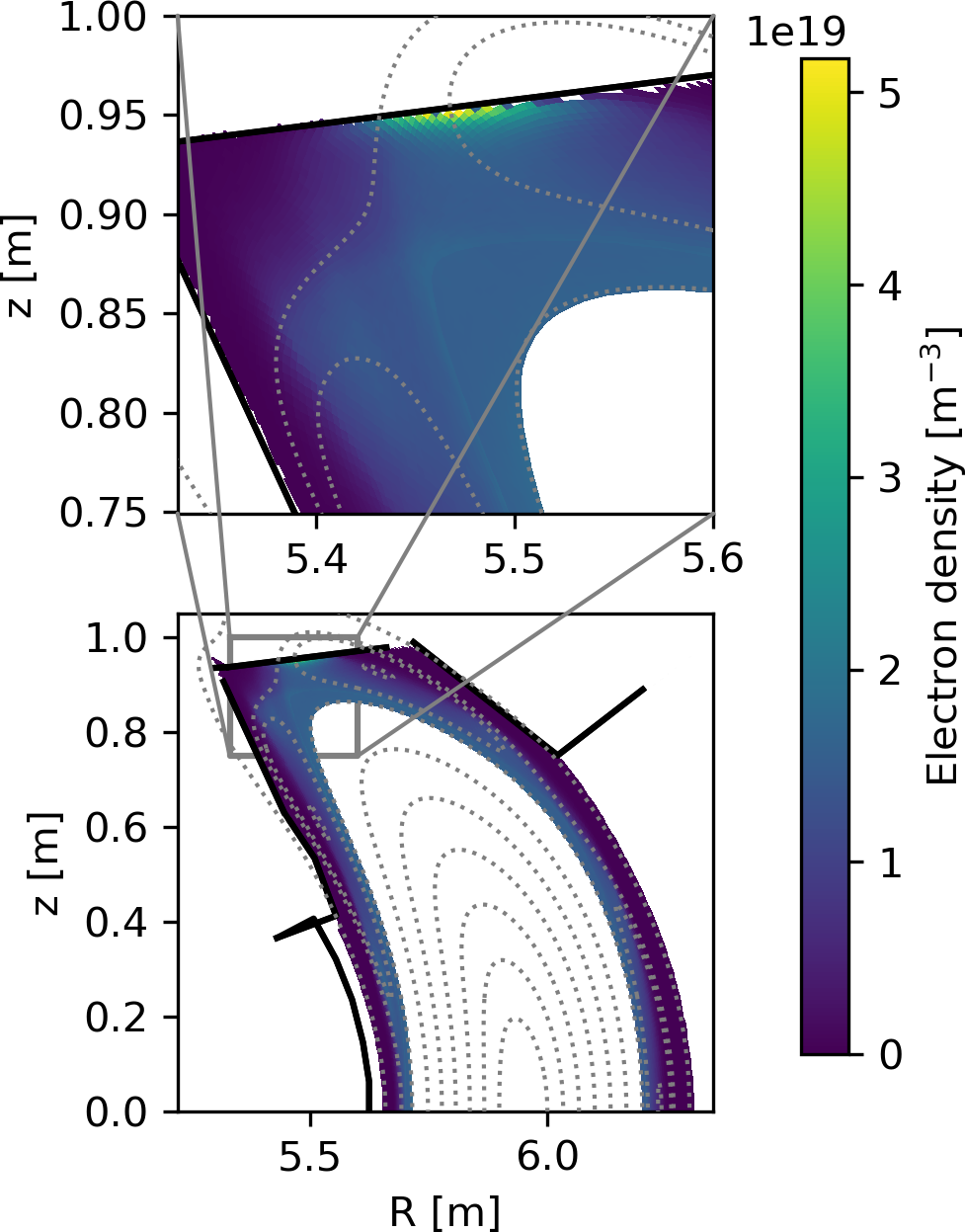}}
  \raisebox{-0.5\height}{\includegraphics[width=.5\linewidth]{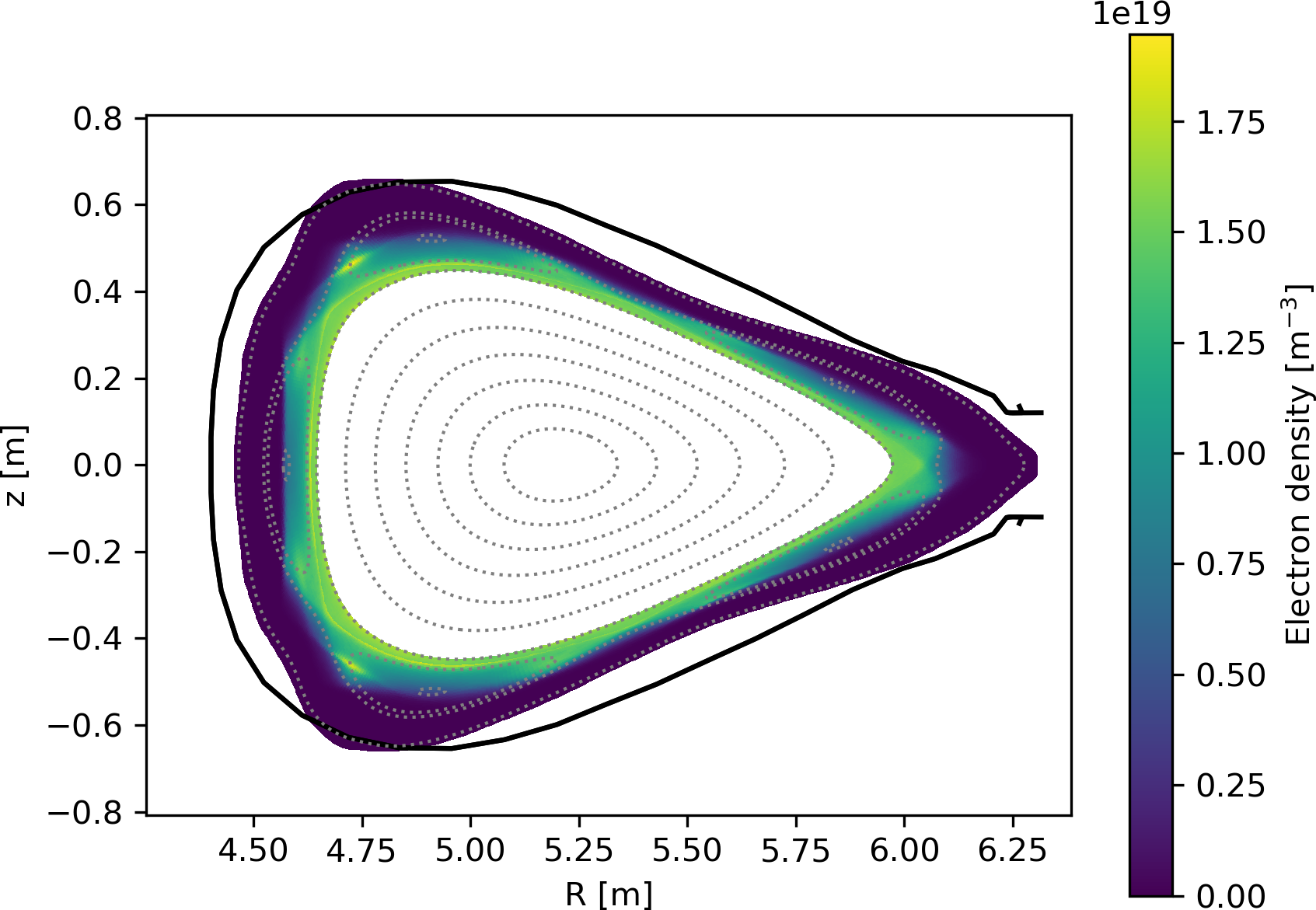}}
  \raisebox{-0.5\height}{\includegraphics[width=.4\linewidth]{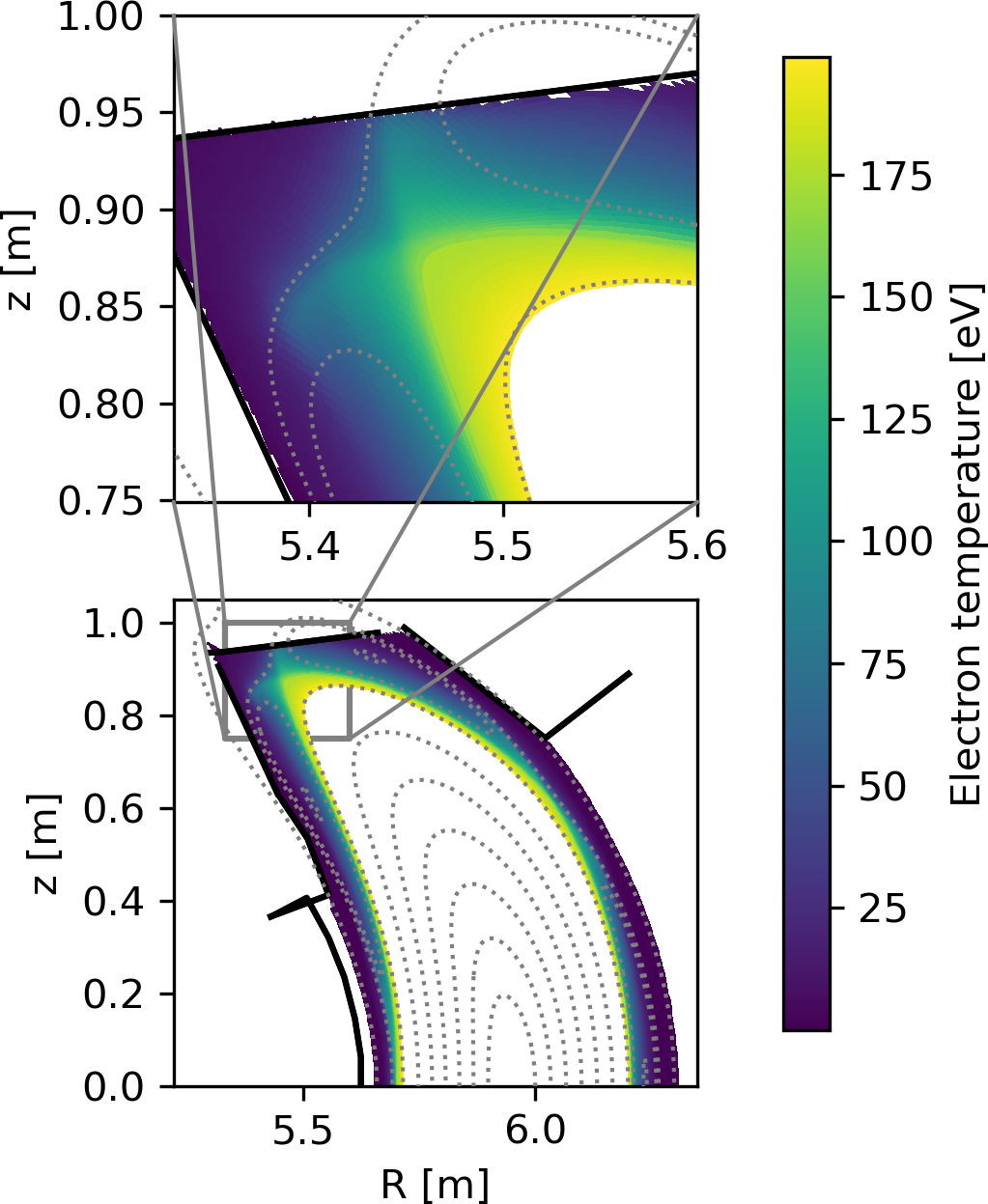}}
  \raisebox{-0.5\height}{\includegraphics[width=.5\linewidth]{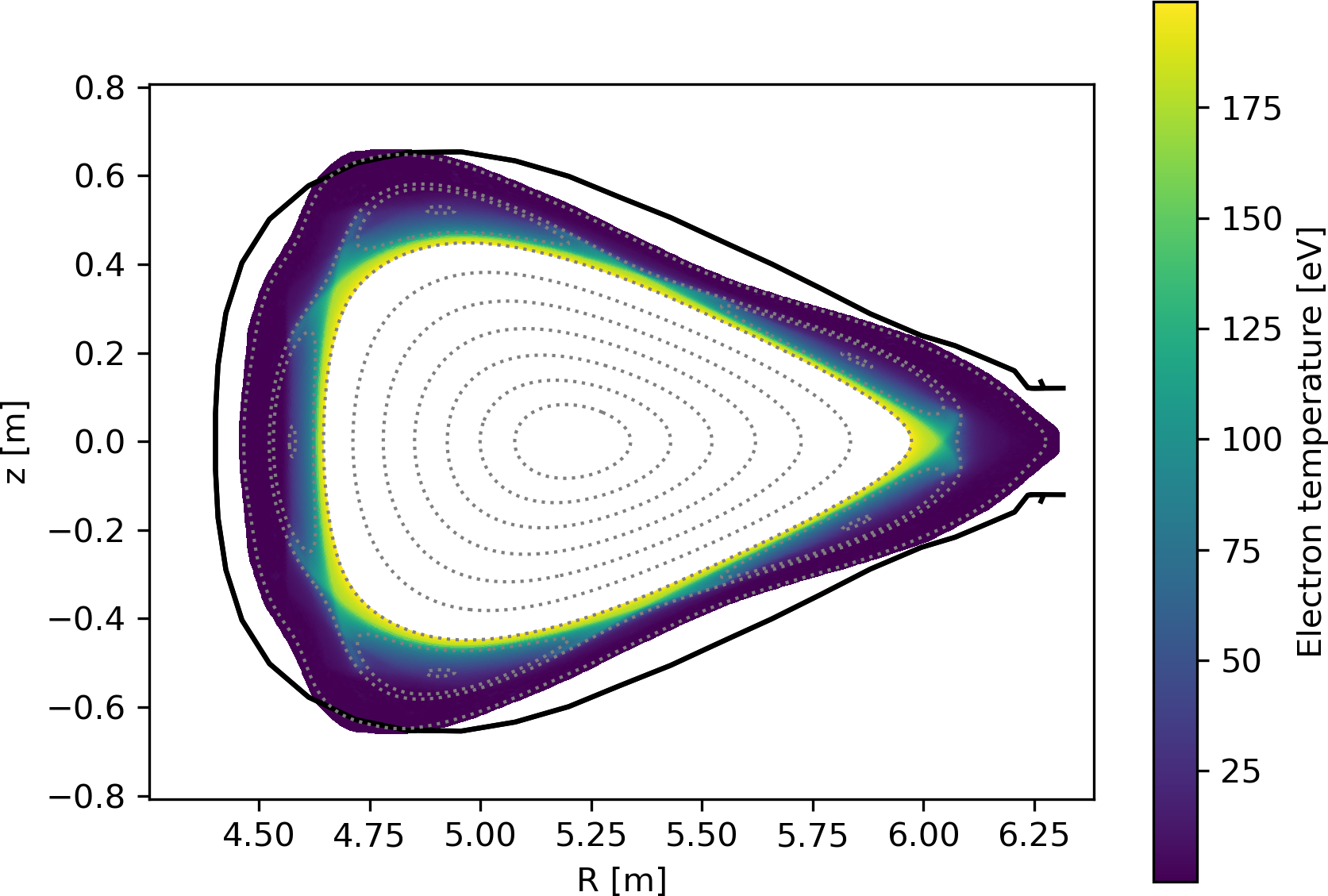}}
  \caption{Density (top) and electron temperature (bottom) profiles for
    scenario~B at
    \dens{1} with \Dis{0.1\ldots 0.4} at the bean shape, at $\varphi=0$ (left) and the
    triangular shape, at $\varphi=\pi / 5 = 36$\,\textdegree{} (right). Shown
    in black are the plasma facing components and as grey dotted lines some
    flux surfaces.
  }\label{f:simnt}
\end{figure}

Fig.~\ref{f:simnt} shows plots of the electron density and temperature
distribution of a simulation, where the diffusion coefficient was
set to \Dis{0.1 \ldots 0.4} from scenario~B and the upstream density was set to
\dens{1}.
The density shows a peak just in front of the target, which is observed in all
simulations. This can be seen at toroidal angle $\varphi=0$, where the upper
and lower target plates are visible.
At the triangular shape
($\varphi=\pi / 5 = 36$\,\textdegree) no target plates are present and, thus, also the density
is not as strongly peaked as in the presence of targets, where recycling is
happening. Note the difference in the colorbar for the density.
In this case the average electron
temperature at the separatrix is around 150\,eV. The average separatrix electron
temperature is for all cases analysed here below 200\,eV.
Experimentally, separatrix electron temperatures were
between 30 and 100\,eV. This deviation will be discussed later in
sec.~\ref{s:mpm}.

\subsection{Toroidal distribution}
\begin{figure}
  \centering
  \includegraphics[width=.73\linewidth, trim={0 4em 0 0.7em}, clip]{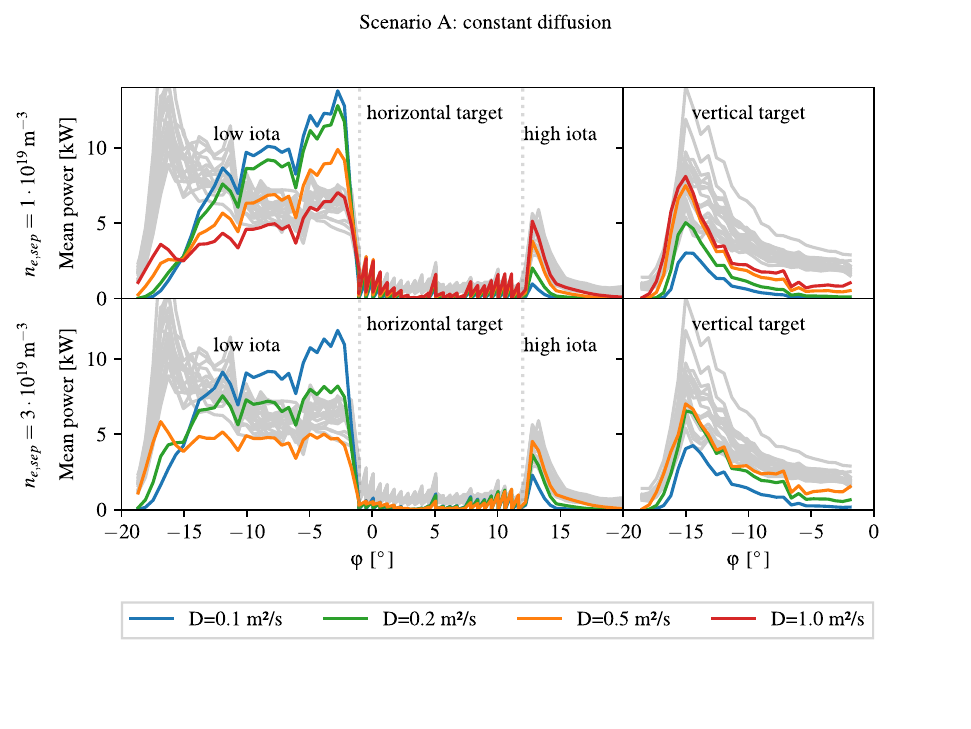}
  \includegraphics[width=.73\linewidth, trim={0 4em 0 0.7em}, clip]{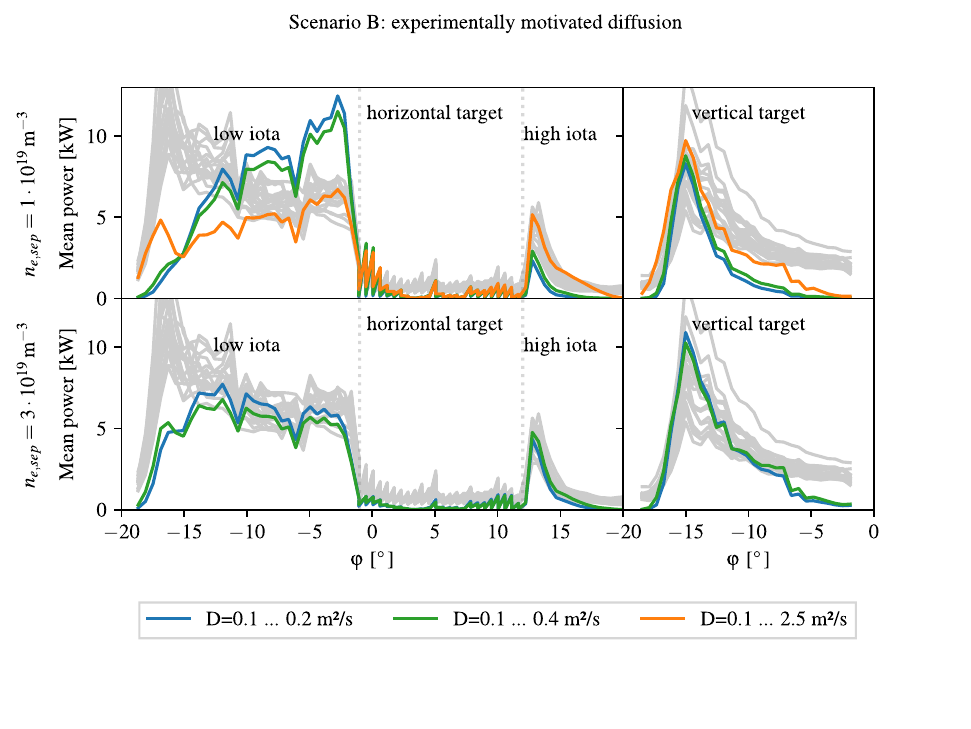}
  \includegraphics[width=.73\linewidth, trim={0 4em 0 0.7em}, clip]{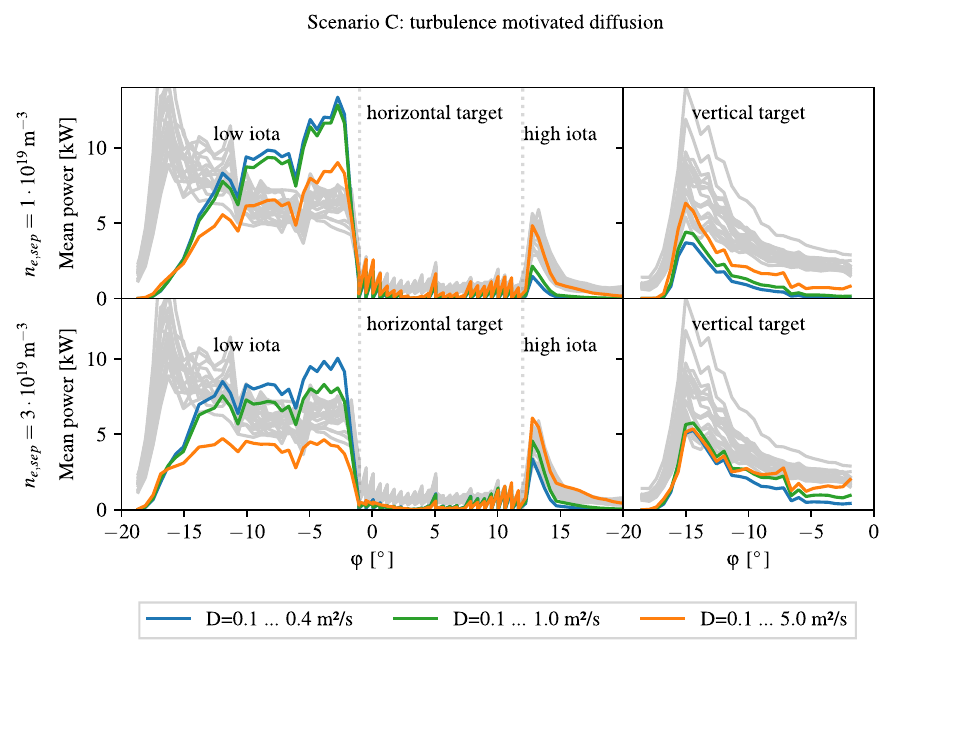}
  \caption{Plot of the power per finger for simulations. On the top are the
    results for scenario~A: constant
    diffusion coefficients.  
    In the middle are the results for scenario~B:
    motivated by experimental observation.
    On the bottom are the results for scenario~C:
    motivated by bad curvature.
    The experimental data is shown in light grey.
  }\label{f:powerfinger}
\end{figure}
The heat-flux on each target finger, as introduced in fig.~\ref{f:fingerpos}, is spatially integrated, giving the toroidal
distribution of the heat-flux. The toroidal distribution for scenario~A is
shown in fig.~\ref{f:powerfinger} on top.
For the non-constant diffusion coefficients the toroidal power
distribution for each density level is
shown in fig.~\ref{f:powerfinger} on the middle and bottom. 
The general trends are similar. For
low diffusion coefficients, with increasing
density less power is deposited on the low iota target (horizontal target at
\phiisless{0}), and more on the high iota target (\phiismore{12}) as well as
the vertical target. The power on
the low iota target is for low density and low
diffusion coefficients peaks at \phiis{-2} and only with
higher density and diffusion a more flat distribution on the low iota target
is observed.
Peaking at \phiis{-17} as in experiments is, in general, not observed
in the simulations. While in the \dens{1} cases the peak is around \phiis{-2}, the
toroidal profiles are more flat for the \dens{3} cases.
In general scenario~B for \dens{3} matches best the experimental results.
The main disagreement is the missing peak at \phiis{-17}.

In all cases in increase in density or diffusion coefficient causes:
\begin{itemize}
\item flattening of the toroidal distribution
\item more power on the vertical target
\item more power on the high iota target
\item less power on the low iota target
\end{itemize}
This is in agreement with
experiments, where the trends are also observed for increasing density.



\subsection{Strike-line}
\begin{figure}
  \centering
  \begin{minipage}{8em}Scenario~A:\end{minipage}
  \includegraphics[width=\linewidth]{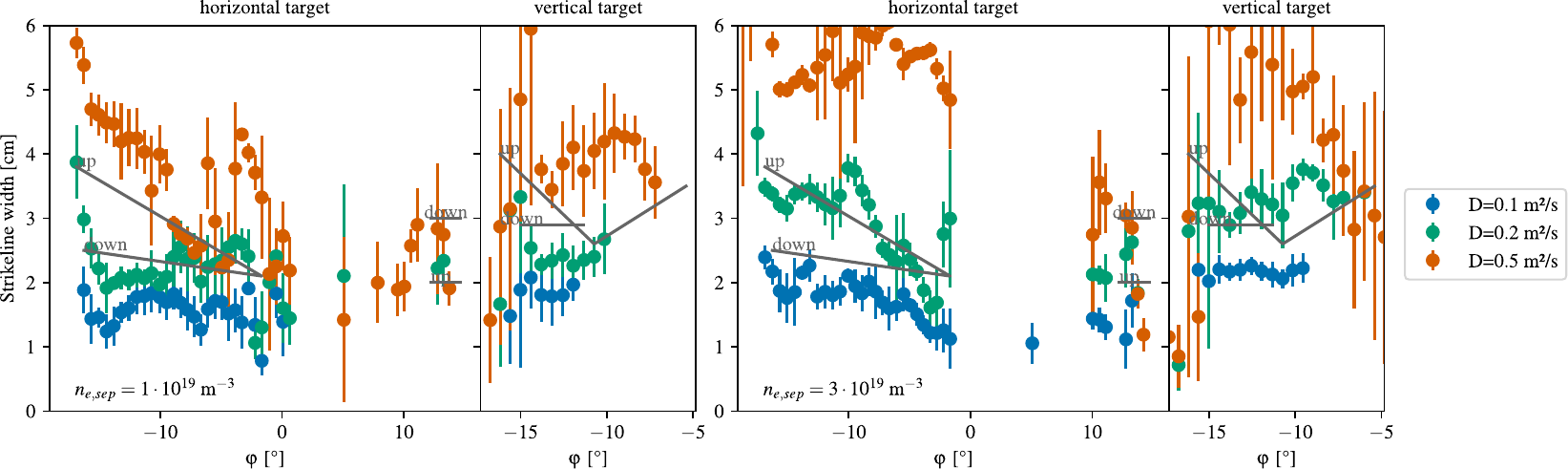}\\
  \begin{minipage}{9em}Scenario~B:\end{minipage}
  \includegraphics[width=\linewidth]{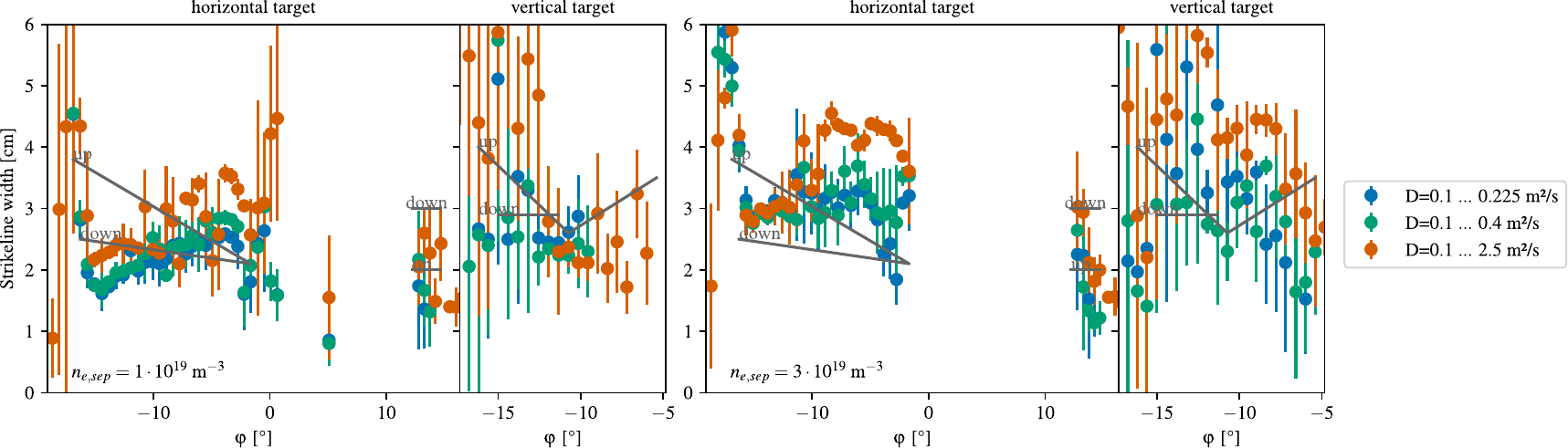}\\
  \begin{minipage}{9em}Scenario~C:\end{minipage}
  \includegraphics[width=\linewidth]{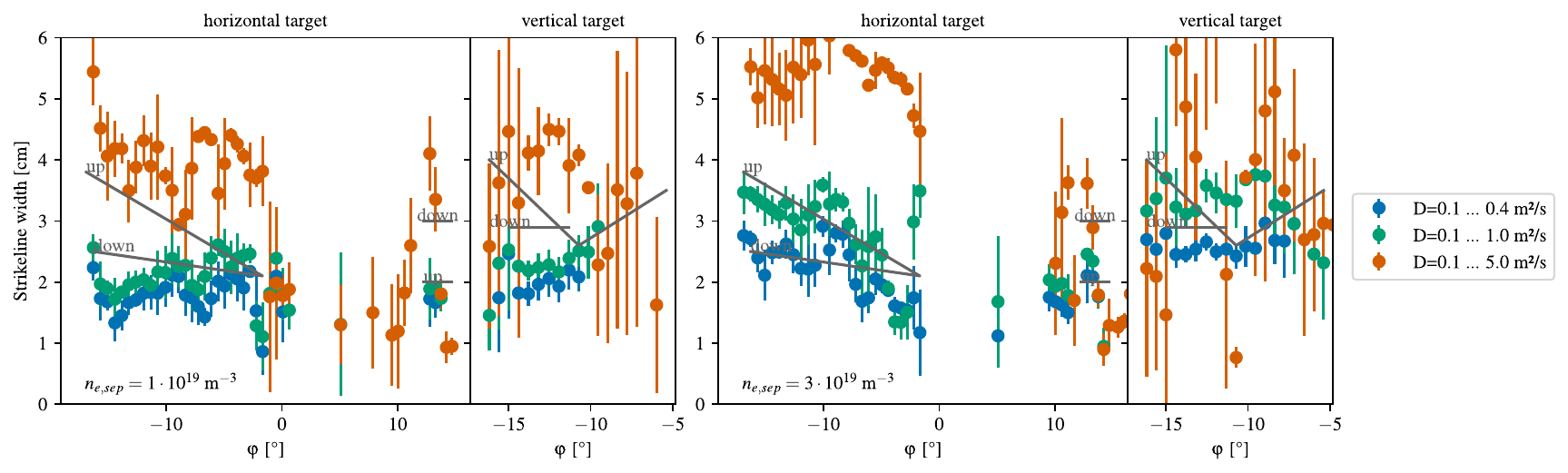}
  \caption{Mean of the strike-line width as a function of the fingers, as
    introduced in fig.~\ref{f:fingerpos}.  The power on the finger is colour
    coded. Simulation results for a constant diffusion \Dis{0.1 \ldots 0.5} on
    top, results for scenario~B introduced in fig.~\ref{f:D:isl} in the middle
    and  results for scenario~C introduced in fig.~\ref{f:D:poltor} on the
    bottom.
    On the left are the results for \dens{1} and on the right for \dens{3}.
    The grey lines show the estimates for the low density case from
    ref.~\cite{bold22a}.
  }\label{f:slw:const}
\end{figure}
The width of the strike
line is of particular interest, as this influences the area over which the heat is
distributed and, thus, also the peak heat-flux that the divertor has to
withstand. Besides this more practical question, the strike-line width gives
also insight into the transport. Fig.~\ref{f:slw:const} shows on top the fitted
strike-line width for a separatrix densities \dens{1} and \dens{3} with a diffusion
coefficient scan in the range
\Dis{0.1\ldots 0.5} for scenario~A: constant diffusion.
The experimentally observed density is
likely between \dens{1} and \dens{3}, as will be later discussed based
on MPM data in sec.~\ref{s:mpm}.

For the lowest \Dis{0.1} the strike-line width is $1\ldots 2$\,cm on the low
iota target, and, thus, smaller than the experimentally observed ones as
indicated by the grey line. The two grey lines are measurements for the upper
and lower divertor. Differences are expected due to drifts.  For \Dis{0.2} the strike-line width is $2\ldots 3$\,cm
matching most closely to the experiment, while for \Dis{0.5} the strike
line width is in the range of $2\ldots 5$\,cm and, thus, a bit larger than
observed.

The peak on the high iota target, on the horizontal target at \phiis{12},
agrees with experiment for \Dop{\ge}{0.5} and \dens{1} spatially constant diffusion
coefficient values.
For the \dens{3} cases, \Dop{\ge}{0.2} matches.




The strike-line widths for the spatially varying diffusion coefficients are shown
in fig.~\ref{f:slw:const} in the middle for scenario~B: diffusion motivated by
experiment and on the bottom
for scenario~C: diffusion motivated by
turbulent transport.
In order to change the strike-line width in the low iota target, a significant
variation of $D$ is needed. In contrast, the strike-line width on the
vertical target is more sensitive to an enhanced transport in the
island then on the horizontal target.
The increased strike-line width coincides with an increased power load on the
vertical target.

From the strike-line width on the low iota target, in scenario~B: motivated by
experimental observations, shown in fig.~\ref{f:slw:const}, the
\Dis{0.1\ldots2.5} case agrees with the experimental observations. For
scenario~C, the experimental strike-line width on the low iota target lies
between the \Dis{0.1\ldots1.0} and the \Dis{0.1\ldots5.0} case.
Thus, for all scenarios we can find cases that give similar strike-line
widths than observed in experiments, all of which feature low level of
\Dapprox{0.1} around the island separatrix.

\subsection{Upstream data}\label{s:mpm}
To further compare the output of the models to experimental data,
upstream data is critical.
As introduced in sec.~\ref{s:diag}, the MPM can measure the density and
temperature in the SOL, outside of the separatrix.
Due to the separation from the targets, this gives a more complete picture,
allowing to (de)validate the transport model.

\begin{figure}
  \centering
  \includegraphics[width=\linewidth]{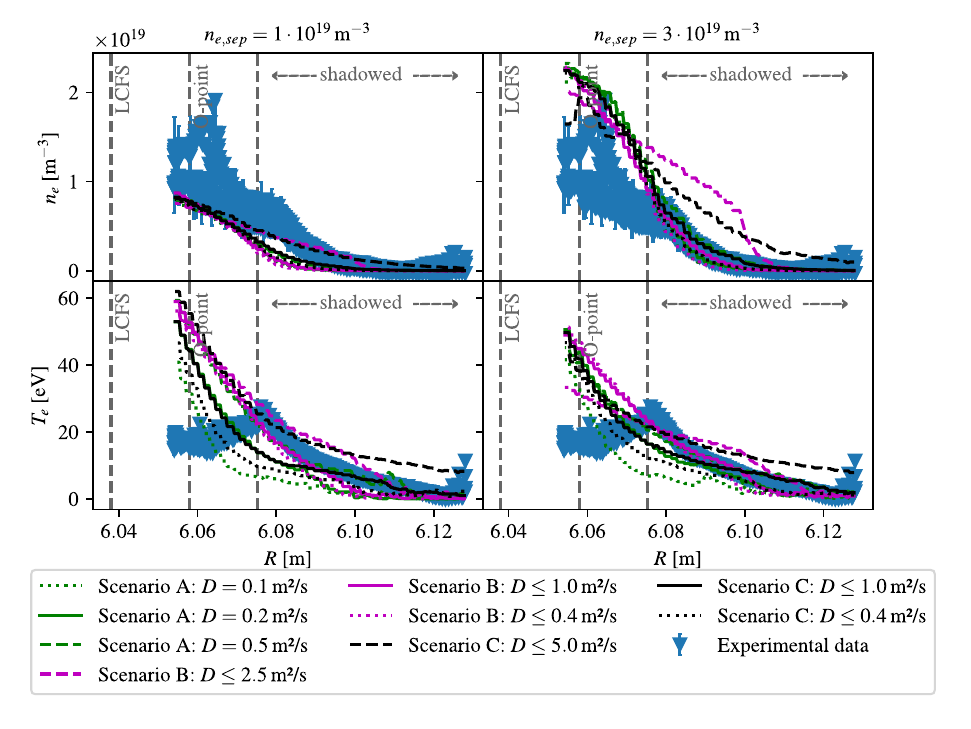}
  \caption{Plot of electron density (top row) and electron temperature (bottom
    row) as a function of
    the radial position. Shown is a 1D cut along the path of the MPM
    diagnostic~\cite{killer21a,killer19a}.  The experimental data from
    the MPM diagnostic is shown as symbols. The differences in the MPM
    measurements are expected as the data is from different
    experiments. Simulation data for \dens{1} is
    shown on the left and for \dens{3} on the right hand side.
    Continuous lines are the
    simulations for the best matching diffusion based on
    strike-line width for the given spatial distribution. Dotted lines
    denote lower diffusion, i.e. more narrow strike-line width and
    dashed lines denote higher diffusion values. For scenario~A:
    constant diffusion \Dis{0.1} is plotted as green dotted, \Dis{0.2}
    as green line (best match based on strike-line width) and \Dis{0.5}
    as green dashed.
    Note that the temperature for \Dis{0.1} is lower in the plotted
    regime, the gradient is higher and the separatrix temperature is
    the highest for the low diffusion value.
    For scenario~B: diffusion motivated by
    experiment, \Dis{0.1 \ldots 0.4} is shown as magenta dotted, \Dis{0.1
      \ldots 2.5} as magenta line (best match based on strike-line width)
    and \Dis{0.1\ldots 10} as magenta dashed. For scenario~C: diffusion
    motivated by turbulence, \Dis{0.1\ldots 1} as black dotted and
    \Dis{0.1\ldots 5} as black line (best match based on strike-line
    width).  The point magnetically closest to the O-point is around
    $R=6.05$\,m while the shadowed area is around $R>6.075$\,m.}
  \label{f:mpm}
\end{figure}

Figure~\ref{f:mpm} shows the density and temperature along the line of sight
of the MPM diagnostic.
Although no experimental MPM data is is available for program
\#20180920.009 and \#20180920.013, similar programs
with MPM data exist and are used as an upstream comparison to
simulation.
The simulation results, for both spatially constant and spatially
varying diffusion coefficients, are plotted as lines in
fig.~\ref{f:mpm}.
The simulations that match the strike-line width best are plotted as
lines, lower $D$ distributions are dotted and higher ones are
dashed, for all three scenarios.

All simulations show essentially monotonic behaviour in the
temperature and the density profiles.
Even though the transport coefficient feature significant changes, the
profiles look relative similar.
This is in contrast to the experimental
data, where the density shows a monotonic, roughly
exponential decay in the shadow region $R > 6.08$\,m but further inside the
gradient vanishes. Towards the centre of the island the density
peaks in some cases, but not in all.
The temperature profiles show less variation.
They show a monotonic trend in the shadowed region
similar to the density.
In the region of longer connection length, towards the O-point of the island,
a hollow temperature profile is observed.  The hollow temperature is
not observed in any of the simulations.

The separatrix density is an input parameter for the simulations. As
such simulations matching best the experimental case can be chosen.
For all diffusion cases, \dens{1} seems to underestimate the density, while
\dens{3} seems to overestimate the density.  Quantifying which case matches
best, or which density
would match best, is not well defined as the profiles do not match qualitatively.
The density for scenario~A, \dens{3} matches very well in the target shadow,
while, depending on the experimental measurement, the \dens{1} case
matches reasonably well towards the O-point.
Scenario~B shows, like the experimental data, a drop in the density
profile but further out around $R\approx 6.10$\,m. This kink is caused by a change in $D$, from the
increased value in the centre of the island to \Dis{0.1} outside the island.
For $R<6.10$\,m the profile is roughly linear, while further
outside an exponential profile is simulated. For the experiments, the
drop is $R\approx 6.08$\,m, with an exponential behaviour outside.

For the temperature, all simulations feature a too high temperature
towards the island centre. Not shown here is the separatrix, where the
temperature in all simulations increases further towards the
separatrix up to 200\,eV, while experimentally values below 100\,eV are
measured.  Scenario~B shows
the lowest separatrix temperature. Due to the large fall-off length,
the temperature is higher further out than observed by the other
simulations.  \Dis{0.1} for scenario~A might appear to show good
agreement, as it features the lowest temperature at $R\approx6.06$\,m, but the
temperature at the separatrix is the highest of all shown simulations.
Similar to the density profile, the temperature profile of
scenario~B shows a kink, featuring a stronger fall off outside.
In general, with increasing $D$ the separatrix temperature decreases,
and the fall-off-length increases.
The impact of the transport coefficient on the simulated profiles will
be analysed and discussed in the following section.



\subsection{Summary}\label{s:sum}
The allow for a stream lined discussion of the findings, a short summary is
given here.
\begin{itemize}
\item Increasing \Dvar{} or $n$ increases the strike-line width. Depending on
  the distribution the dependency is weak.
\item Increasing \Dvar{} or $n$ changes the toroidal heat-flux distribution. In
  general the distribution on the low iota target gets less peaked, and more power
  is reaching the high iota and vertical target.
\item There is a mismatch in the target shadow region, e.g. on the horizontal
  target around \phiis{-17} and \phirange{0}{10} and on the
  vertical target around \phiis{-17} as well as \phiis{-5}.
\item In the experiment the target heat flux peaks around \phiis{-17} and
  decreases towards \phiis{-2}, while in the simulations it peaks at
  \phiis{-2} and decreases towards \phiis{-17}. As the distribution becomes
  more flat with higher \Dvar{} or $n$, the difference reduces.
\item The upstream measurements show a vanishing gradiant at the point closest
  to the O-point, the simulations show in all cases a monotonic decrease with
  radius.
\item The monotonic behaviour in the radial profiles is resilient to the
  tested variation in the transport coefficients.
\item The upstream temperature observed in simulations is in all cases too
  high, even for higher density cases.
\end{itemize}

\section{Discussion}\label{s:discussion}
In this section the results from sec.~\ref{s:simulations} will be discussed.
Note that a short summary of the findings is given in sec.~\ref{s:sum}.



\subsection{Transport models}
Besides scenario~A: featuring constant diffusion, two different spatially
varying diffusion coefficient distributions
have been tested in EMC3-Eirene.
The three scenarios have been used for a scan in both density as well
as the magnitude of the diffusion coefficient, using the EMC3-Eirene code.
A one-to-one comparison of the heat-flux from experiment and
simulations were performed.

\subsubsection{Scenario~B: experimentally motivated}
\begin{figure}
  \centering
  \includegraphics[width=.5\linewidth]{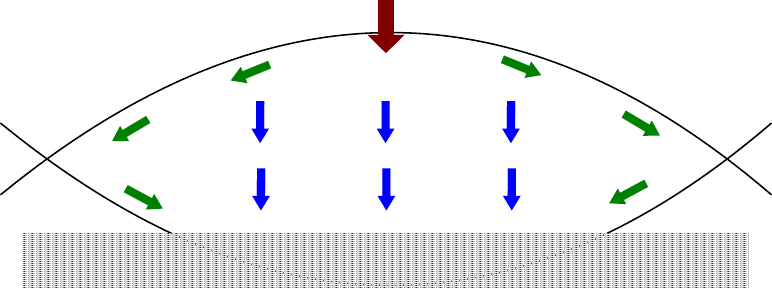}
  \caption{Sketch of the heat-flux in the SOL of \mbox{W7-X}. Depicted is the cross
    section of an island. The target is on the bottom. The heat-flux $P_s$ is
    entering at the top (red arrow).
    The channel via parallel transport $P_\parallel$ at the edge of the island is shown as green
    arrows, while the transport through the bulk of the island $P_\perp$, relying on diffusion, is
    shown as blue arrows.}
  \label{f:2chan}
\end{figure}
Figure~\ref{f:2chan} sketches the simplified transport through the island in
the island divertor on a single island.  The heat from the confined region $P_s$
enters the island through the separatrix, which separates the island from the
confined region, depicted as a red arrow.  This is not localised,
but spread over the entire separatrix.
From the region where the heat enters the island, there are,
in a simple picture, two channels to the target: the parallel channel and the
perpendicular channel. The parallel channel is depicted by green arrows in fig.~\ref{f:2chan}.
In tokamaks this is the
main channel of transport in attached conditions and can be described by the
two-point-model~\cite{borrass91a,pitcher97a,schworer20b}.  The heat-flux is
described by the Spitzer conduction, scaling with temperature
$q\propto T^{5/2}$, that is, a strong function of temperature.
As $T$ drops radially into the island the parallel transport
becomes less efficient.
Thus, only a width of the order of the temperature
decay length $\lambda_T$ contributes to the parallel channel.  The width in turn
is determined by the perpendicular heat transport $\lambda_T\propto\chi_s$ at
the island edge.
Note, that in a more complete model, $\chi_s$ needs to be replaced by
a flux-tube averaged quantity.
This gives a total power going through the parallel channel as
$P_\parallel \propto \chi_s \times T_s^{5/2}$, with $T_s$ the separatrix
temperature. The transport through the bulk of the island, depicted in blue, in turn
depends on the temperature at the separatrix.
After the initial $\lambda_T$, a change in $\chi$ does not
directly affect the parallel heat-flux.
Indirect effects do exist:
For a given separatrix heat-flux $P_s$, $P_\parallel$ needs to
decrease if $P_\perp$ increases, resulting in a lower separatrix
temperature $T_s$.
These indirect can be significant.
The heat-flux for a single perpendicular
channel is proportional to the ``integrated heat conductivity'':
$X = 1/\int_{sep}^{tar} \frac{1}{\chi} dr$. It is also proportional to
the temperature difference between the temperature $\lambda_q$ from
the separatrix and the temperature at the PFC $T\approx 0$.
Thus the temperature difference is 
$T(R_{sep}+\lambda_T) - 0 = T_s/e \propto T$.
The total heat-flux through the bulk of the island $P_\perp$ is given by the sum over all
possible channels $P_\perp \propto T_s \iint X dA$.
Thus increasing $\chi$ in the bulk of the island, increases $X$ and, thus, $P_\perp$
without a direct effect on $\lambda_T$. This allows to retain the peak shape,
but lower upstream temperature, as power is put into the far SOL.
This is seen in the results for scenario~B, where the enhanced
transport towards the centre of the island retains a narrow SLW, but
allows to decrease the separatrix temperature.

\subsubsection{Scenario~C: turbulence scaling}
Scenario~C is based on the scaling of turbulent transport with bad
curvature, which is in \mbox{W7-X} largest at the outer bean cross-section.
Similar to scenario~B, this scenario allows us to decrease the separatrix
temperature, without a significant impact on the SLW.
This is in contrast to scenario~A, where the decoupling was not
possible.
The transport is enhanced at the edge of the separatrix, but not at,
or below,
the x-points, as shown in fig.~\ref{f:D:poltor}. The main contribution to the
parallel connection length is around the x-point.
This could be an explanation why scenario~C features a reduced upstream
temperature and a narrow strike-line, increasing agreement with
experimental measurements.

\subsubsection{Distinguishability of scenarios}
The strike-line width from the experiments has been determined to be around 2\,cm
to 4$\,$cm in the magnetic standard configuration. For all scenarios and
densities, cases have been found that are consistent with the magnitude of
the strike-line. The observed SLW of 2\,cm to 4\,cm is true for the low
iota target, the high iota target as well as the vertical target, independent
of the connection length.  This observation is reproduced in the simulations,
where for a given density and diffusion the strike-line width is roughly
constant for all significantly loaded areas.
As also the toroidal distribution features similar shape for all
scenarios, it is not clear whether it is possible to distinguish the appropriate transport
model based purely on the target profiles.
The different scenarios show clearly different radial profiles at the
line of sight of the MPM.
For the density profiles, scenario~B and scenario~C show for large $D$
values an increased density in the shadowed volume. Scenario~B drops
of sharpely at the point where $D$ is decreased again, while
scenario~C does not feature this drop, as the diffusion coefficient
does not has a radial dependency.
For the temperature profiles, more variation is observed.
The temperature falloff length varies significantly between the
different simulations. The shortest falloff length is observed for
scenario~A \Dis{0.1} while the largest falloff length is observed for
scenario~B \Dis{0.1\ldots 2.5}.
None of the cases show good agreement with the measured
profiles, as the experimental data shows vanishing gradients closest
to the O-point, while the simulations all show a monotonic behaviour.

The inclusion of upstream measurements, like from the MPM, is crucial
for accurately comparing transport models and experiments. This	is due
to the observed ``insensitivity'' of the target profiles to the different
scenarios, where for all scenarios a good match with experiments has
been found.

\Dis{0.2} and \dens{1} from scenario~A matches the lower
divertors. The upper divertor is better matched by an increased
diffusion, case \Dis{0.5} and \dens{1}, or increased density, case
\dens{3} and \Dis{0.2}.
The up-down asymmetry reverses if the magnetic field is
reversed~\cite{zhang21a}, in line with the effects of drifts.

\subsection{Drifts}
\begin{figure}
  \centering
  \includegraphics[width=.9\linewidth]{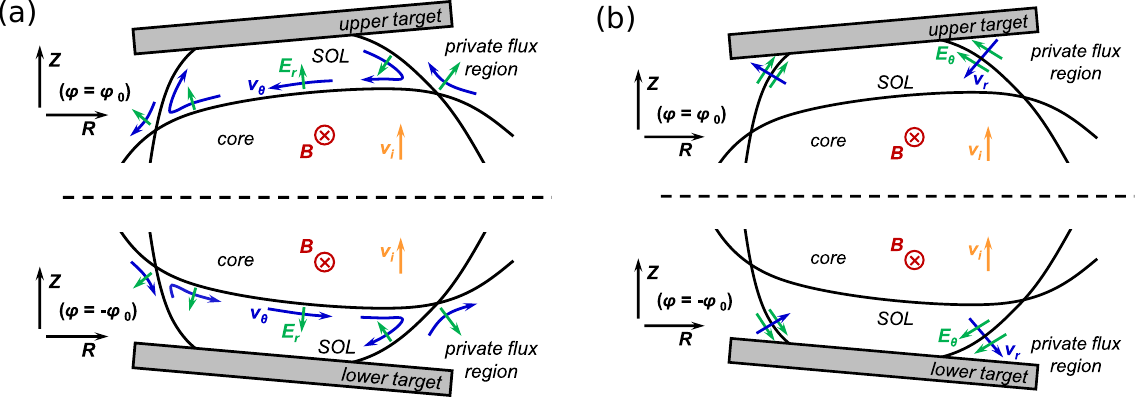}
  \caption{Sketch of the $\vec E \times \vec B$ drifts in \mbox{W7-X}.
    Based on the work of Hammond\etaln~\cite{hammond19a}.}\label{f:drift}
\end{figure}
Field errors could cause the variations between half modules, but the
variation between upper versus lower half modules seems rather
systematic, even after symmetrisation with error field
correction~\cite{bold22a}.
Drifts are expected to cause up-down asymmetries as
depicted in fig.~\ref{f:drift}~\cite{hammond19a}.
Experiments with reversed field can be used to test this hypothesis.
Extending this analysis to low or high iota cases would also be interresting,
as in those cases the impact of error fields is expected to reduce. However,
due to the changed SOL geometry, the effect of drifts is also likely to
change.

The location of the strike-line, shown in ref.~\cite{bold22a}, is in
agreement with Hammond\etaln~\cite{hammond19a} where, in the low iota
forward configuration, the peak heat load on the low iota target was a few
cm closer to the pumping gap compared to the magnetic strike line,
which was attributed to drifts.

%
Experiments of field reversal in low iota
configuration~\cite{hammond19a} and standard configuration~\cite{zhang21a} imply drifts
are responsible for up-down asymmetries.
The experiments include error field correction, i.e. the magnetic field has
been optimised to minimise differences between the different target heat
loads~\cite{lazerson18a} and should thus be of reduced importance,
leaving an exacerbated role for drifts.
Besides drifts, a systematic
misalignment of the divertors or vertical misplacement of the entire
plasma could explain the up-down assymmetry.

\subsubsection{Impact of drifts on SLW}
The strike-line width varied between upper and
lower divertor~\cite{bold22a}.
If one assumes that this is based on drifts, one can make a rough
estimate on the effective parallel drift velocity using a simple
model: The parallel velocity is assumed to be the sound speed $c_s$,
modified by the drift velocity, with the sign of the drift depending
on upper or lower divertors. Rather then the actual drift velocity, the
effective parallel drift velocity $v_{D,\parallel}$.
The projection of $v_{D,\parallel}$ on the perpendicular plane gives the drift
velocity $v_{D,\perp}$.
The estimated parallel transport time is then based on the parallel
connection length $L_\parallel$ and given by
$L_\parallel / (c_s \pm )$.
This gives a strike-line width of around
\begin{align}
  \lambda_\pm &\approx \sqrt{\frac{\chi L_\parallel}{c_s \pm v_{D,\parallel}}}
  \intertext{which in turn gives an effective drift speed of}
  v_{D,\parallel} &\approx \frac{\chi L_\parallel}{2}
  \frac{\lambda^2_--\lambda^2_+}{\lambda^2_+\lambda^2_-}
  \intertext{or an estimate of the heat diffusion $\chi$}
  \chi &\approx \frac{2 c_s}{L}
  \frac{\lambda^2_+\lambda^2_-}{\lambda^2_++\lambda^2_-}
\end{align}
Using $\lambda_- = 3\,$cm and $\lambda_+ = 2\,$cm, and assuming
$L_\parallel \approx 100\,$m and $\chi = 0.3\,$m$^2/s$ this gives an
effective parallel contribution of the drifts on the order of
$v_{D,\parallel} \approx 21\,$km/s and, thus, in the range of the
speed of sound, which is around 30\,km/s.
Using the velocity of 30\,km/s gives \chiapprox{0.17} and thus in the range
that was used in the simulations.

As the field line pitch in W7-X is around 0.001, means that $v_{D,\perp}$ is
around 1000 times smaller then $v_{D,\parallel}$.
Thus $v_{D,\perp}\approx 21$\,m/s and thus several orders of magnitude smaller
then experimentally measured~\cite{flom23a}.
Thus it seems this simple SOL model for the impact of drifts on not valid.
Reasons might be that the dwell time in the SOL is not set by the parallel
dynamic, but rather by perpendicular process. This would be in contrast to
tokamaks, where the field line pitch is much smaller.

\subsubsection{Toroidal distribution}
While it is possible to match the strike-line width to the experimental ones,
none of the simulations matched the toroidal distribution on the low iota
target. The flattening with higher $n$ and $D$ is however observed in all cases.
The toroidal distribution does not appear to be strongly influenced by the
diffusion model chosen.
There is a weak dependency on density and diffusion
magnitude, which has already been discussed for
scenario~A~\cite{bold22a}.

Scenario~B with \dens{3} shows, compared to the other simulation, an increased
heat-flux at \phiis{-17}. However, a strong peaking, as in the experiments at
\phiis{-17} is not reproduced.
As those field lines ending at
\phiis{-17} are in the private flux region (PFR), it requires
additional perpendicular transport. That could be anomalous or drift driven.
A toroidal redistribution of fluxes could be
associated with $E \times B$ drifts. That would require poloidal electric
fields, that cause a radial drift into the PFR, as shown in
fig.~\ref{f:drift}.

\subsubsection{Hollow island}
The hollow temperature profile, as measured by the MPM in
the island has not been reproduced by any of the simulations.
In order to do so, a significant sink in the centre, e.g. due to
radiation, or a significantly transport reduction into the centre is
needed.
%
The density
profiles measured by the MPM agree better with the non-constant
diffusion coefficients, selected from the simulation with a matching strike
line width.
The hollow temperature profile and the associated non-monotonic density
profile in the island have been repeatedly experimentally observed in the
past also by other diagnostics in different
locations~\cite{killer19a,killer19b,drews19a,killer21a,killer21b,barbui19a,flom23a}.
In EMC3-Eirene simulations, this has been reproduced by locally
reducing the heat transport in the centre of the island. To reproduce a
hollow temperature profile, the heat diffusion coefficient in the
island needs to be reduced by an order of magnitude, and in order to
get similar results to experiments by two orders of magnitude.
It is not obvious why heat transport in the island centre should be that
strongly surpressed.
The lack of match in the here presented simulations shows that the
current model are either missing a fast transport channel around the
island centre, or a significant transport reduction into the island.
The fast transport around the island would connect a region outside of
the O-point to a point inside. This would allow to transport the heat
around the island, without the heat going through the island centre.
To help elucidate how important drifts may be, we assume the hollow island is
caused by fast convective transport around the island centre, and drifts are
solely responsible for the convective transport.
An estimate of the velocity can then be
given. This assumes that the transport time is given by the square of
the fall-off-length $\lambda_T$ over the diffusion
coefficient. For the poloidal convective
transport to be sufficiently fast to compete with the diffusive inward transport,
the poloidal transport needs to fulfil:
\begin{align}
  v_{pol} &\gtrapprox L_{pol} \chi \left(\frac{\nabla T}{T}\right)^2 \approx
  L_{pol} \chi \frac 1 {\lambda_T^{2}}
\end{align}
Using a falloff length of $\lambda_T\approx 3\,$cm, consistent with the
data presented here, the poloidal island size $L_{pol}$ on the order
of one meter, and $\chi \approx 0.3\,$m$^2$/s gives
$v_{pol} \gtrapprox 300\,$m/s. Parallel flow velocities observed by experiments and
EMC3-Eirene are typically in the range of 30\,km/s. The field line
pitch is around 0.001, giving a perpendicular contribution of
300\,m/s. This suggests that drifts could be comparable to parallel flows,
which is in
line with the above discussion. The estimate of $\gtrapprox 300\,$m/s
is however larger then the above estimated $\approx 21\,$m/s for the
perpendicular velocity.
The drift velocity measured by the MPM are consistent with this
estimate~\cite{killer19b}.
This can also be incorporated into the two-channel model shown in
fig.~\ref{f:2chan}. Assuming a fast enough drift velocity within the
island but outside of the power carrying layer, $\chi$ can be
increased ($\to \infty$) by the part of the path that is governed by fast
poloidal rotation and, thus, giving an effectively increased $X$.
The motivation for such an assumption is, that if the drift is fast
enough, this results in a constant temperature within the closed
poloidal loop.

\section{Conclusion and Summary}\label{s:conclusion}
Please note that a short summary of the results is already given in
sec.~\ref{s:sum}. As such only results are included, that are needed for the
discussion.


%
%
The quantitative comparison performed
here shows that, in order to reproduce the experimentally observed strike
line width in the range of 2 to 4\,cm, diffusion coefficients of
0.2\,m$^2/$s around the separatrix are needed in the magnetic standard configuration for
low to medium density cases.  In addition to constant diffusion coefficients,
spatially varying diffusion coefficients can be used to reproduce the
experimentally observed strike-line width.
The main features at the target, namely SLW and toroidal distribution, were
not strongly sensitive to the chosen transport model at moderate values of
$D$. 
This allows to tune the magnitude of the spatially varying diffusion
coefficients to match other quantities, in this case to reduce the
separatrix temperature.


There are significant differences between simulations and
experiments, that could not be reproduced by the simulations, even
using strongly spatially varying diffusion coefficients.
Some differences are expected to be due to, or seem to be consistent with, a lack of drifts in
EMC3-Eirene: e.g. the
up-down asymmetry on the divertor target plates, the hollow island or
the to high upstream temperature.
Other discrepancies
remain which are not expected, such as the difference in
the toroidal distribution of the heat-flux which peaks at \phiis{-17}
on the low iota target.
Additionally, hollow temperature profiles in the islands are measured by
probes and other diagnostics, which has not been reproduced by the
simulations. Non-isotropic
transport in the perpendicular plane could help to reproduce this, but
convective perpendicular transport due to drifts would also be in
agreement with the experimental measurements.

Altogether, these observations show that
spatially varying diffusion coefficients can improve agreement to experimental
measurements, but so far the applied variations have not been able to
match experimental measurements consistently up and downstream.
The observed discrepancies and experimental measurements further indicate that drifts could contribute
substantially to the transport in \mbox{W7-X}, which cannot be captured with
the present version of EMC3-Eirene. Thus, implementation of the drifts
are a priority for future work in order to fully capture the physics seen in
experiments and determine if they are sufficient to explain the discussed
discrepancies.

\section{Acknowledgement}
This work has been carried out using the xarray
framework~\cite{hoyer17a,xarray_0_17_0}.
Some task have been paralellised using GNU parallel~\cite{tange18a}.

The simulation presented here are available at:
DOI: 10.5281/zenodo.5762079

This work has been carried out within the framework of the EUROfusion
Consortium, funded by the European Union via the Euratom Research and Training
Programme (Grant Agreement No 101052200--EUROfusion). Views and opinions
expressed are however those of the author(s) only and do not necessarily
reflect those of the European Union or the European Commission. Neither the
European Union nor the European Commission can be held responsible for them.

\bibliographystyle{iaea_misc}
\bibliography{phd}

\end{document}